\documentclass[a4paper,11pt]{article}
\usepackage{jheppub}

\usepackage{bm}
\usepackage{color}
\usepackage{amssymb}
\usepackage{amsmath}

\usepackage{MnSymbol}
\usepackage{multirow}
\usepackage{diagbox}

\usepackage{epsfig}

\usepackage{subcaption}
\newcommand{\bs}[1]{\boldsymbol{#1}}

\begin{document}

\title{Gluonic nucleon energy correlators and fracture functions for Color Glass Condensate}

\author[a,b]{Heikki Mäntysaari,}
\emailAdd{heikki.mantysaari@jyu.fi}

\author[c,d]{Yu Shi,}
\emailAdd{yu.shi@polytechnique.edu}

\author[e]{Yossathorn Tawabutr,}
\emailAdd{yossathorn.t@chula.ac.th}

\author[a,b]{Xuan-Bo Tong}\emailAdd{xuan.bo.tong@jyu.fi}

\affiliation[a]{Department of Physics, University of Jyväskylä, P.O. Box 35, 40014 University of Jyväskylä, Finland}
\affiliation[b]{Helsinki Institute of Physics, P.O. Box 64, 00014 University of Helsinki, Finland}

\affiliation[c]{Key Laboratory of Particle Physics and Particle Irradiation (MOE), Institute of frontier and interdisciplinary science, Shandong University, Qingdao, Shandong 266237, China}

\affiliation[d]{CPHT, CNRS, \'Ecole polytechnique,  Institut Polytechnique de Paris, 91120 Palaiseau, France}

\affiliation[e]{Center of Excellence in High Energy Physics, Faculty of Science, Chulalongkorn University,
         254 Phaya Thai Rd, Wang Mai, Pathum Wan,
         Bangkok 10330, Thailand}

\abstract{Nucleon energy correlators (NECs) and fracture functions provide novel tools for probing nucleon and nuclear structure at small $x$ through measurements in the target fragmentation region (TFR) of deep inelastic scattering (DIS). We investigate gluonic NECs and fracture functions using the Color Glass Condensate effective theory at eikonal accuracy. We find that only the unpolarized and linearly polarized gluon components in an unpolarized target are nonvanishing at this order, and that both are determined by the adjoint dipole $S$-matrix. Furthermore, we show that the linearly polarized gluonic NEC $h_{1}^{t,g}$ generates a characteristic $\cos 2\phi$ azimuthal asymmetry in the DIS energy pattern in the TFR. Unlike analogous observables in the current fragmentation region, this asymmetry is governed by the ratio of the {\it linearly polarized gluon} NEC $h_{1}^{t,g}$ to the {\it unpolarized quark} NEC $f_{1}^{q}$, making it particularly sensitive to the saturation scale. Our numerical analysis shows that this asymmetry exhibits substantial nuclear suppression, providing a novel window into the onset of gluon saturation at the future Electron-Ion Collider.
}

\maketitle
\flushbottom

\section{Introduction}

Understanding the small-$x$ gluon structure of nucleons and nuclei is a central goal of high-energy quantum chromodynamics (QCD) and a major scientific objective of the future Electron--Ion Collider (EIC)~\cite{Accardi:2012qut,Aschenauer:2017jsk,AbdulKhalek:2021gbh,Abir:2023fpo}. This program encompasses  the search for parton saturation driven by nonlinear QCD dynamics as well as the tomography of the spatial, momentum, and spin structure of small-$x$ partons in hadronic and nuclear targets. The relevant high-energy dynamics are systematically described by the Color Glass Condensate (CGC) effective theory~\cite{Iancu:2003xm,Gelis:2010nm,Kovchegov:2012mbw,Morreale:2021pnn}, in which Wilson-line correlators constitute the fundamental quantities encoding the multiple scattering of energetic probes off dense gluonic matter.

Substantial progress has been made in relating CGC Wilson-line correlators to conventional parton distributions used in nucleon tomography~(see e.g.,~\cite{Marquet:2009ca,Dominguez:2010xd,Dominguez:2011wm,Xiao:2017yya,Kovchegov:2015pbl,Kovchegov:2018znm,Kovchegov:2018zeq,Cougoulic:2022gbk,Adamiak:2024khm,Caucal:2025xxh,Hatta:2016dxp,Hatta:2017cte,Boer:2018vdi,Bhattacharya:2024sno,Kovchegov:2025yyl,Bhattacharya:2025fnz,Kovchegov:2026gwb,Benic:2026idy,Fu:2023jqv,Hagiwara:2024wqz}). Much of this effort has focused on transverse-momentum-dependent (TMD) parton distributions~\cite{Boussarie:2023izj}, which characterize correlations involving the transverse momentum of an initial-state parton. A broad class of TMDs, including Sivers function~\cite{Boer:2015pni,Dong:2018wsp,Yao:2018vcg,Kovchegov:2021iyc,Kovchegov:2022kyy} and linearly polarized gluon distributions~\cite{Metz:2011wb,Dominguez:2011br,Dumitru:2015gaa,Boer:2016fqd}, has been studied in the CGC framework~\cite{Marquet:2009ca,Dominguez:2010xd,Dominguez:2011wm,Xiao:2017yya,Kovchegov:2015pbl,Kovchegov:2018znm,Kovchegov:2018zeq,Cougoulic:2022gbk,Adamiak:2024khm,Caucal:2025xxh,Mueller:2013wwa,
Boer:2015pni,Dong:2018wsp,Yao:2018vcg,Kovchegov:2021iyc,Kovchegov:2022kyy,
Metz:2011wb,Dominguez:2011br,Dumitru:2015gaa,Boer:2017xpy,Boer:2016fqd}. These studies provide a unified description of semi-inclusive particle production at small transverse momentum in high-energy $eA$ and $pA$ collisions~\cite{
Dominguez:2010xd,Dominguez:2011wm,
Marquet:2009ca,Xiao:2010sa,Xiao:2010sp,Caucal:2024vbv,Altinoluk:2024vgg,Altinoluk:2025ang,Caucal:2025xxh,
Metz:2011wb,Dominguez:2011br,Dumitru:2015gaa,Dumitru:2016jku,Dumitru:2018kuw,Boer:2016fqd,Boussarie:2021ybe,Caucal:2022ulg,Caucal:2023fsf,Altinoluk:2024zom,Caucal:2025mth,Mukherjee:2026cte,Mukherjee:2026six,
Tong:2022zwp,Tong:2023bus,
Mueller:2013wwa,
Stasto:2011ru,Marquet:2016cgx,Boer:2017xpy,Marquet:2017xwy,Stasto:2018rci,Albacete:2018ruq,Marquet:2019ltn,Altinoluk:2023hfz,Altinoluk:2024tyx,Marquet:2025jdr,Caucal:2025zkl, Gao:2026azd}. In deep-inelastic scattering (DIS), these developments have been extensively applied to observables in the current fragmentation region (CFR)~\cite{Boglione:2016bph}, including single-hadron (or jet) production~\cite{Marquet:2009ca,Xiao:2010sa,Xiao:2010sp,Caucal:2024vbv,Altinoluk:2024vgg,Altinoluk:2025ang,Caucal:2025xxh}, dihadron (or dijet) correlations~\cite{Dominguez:2011wm,Mueller:2013wwa,Metz:2011wb,Dominguez:2011br,Dumitru:2015gaa,Dumitru:2016jku,Dumitru:2018kuw,Boer:2016fqd,Boussarie:2021ybe,Caucal:2022ulg,Caucal:2023fsf,Altinoluk:2024zom,Caucal:2025mth,Mukherjee:2026cte,Mukherjee:2026six}, as well as lepton-jet correlations~\cite{Tong:2022zwp,Tong:2023bus}.

\vspace{0.2cm}
\begin{figure}[htbp]
    \centering
        \captionsetup[subfigure]{skip=10pt}

    \begin{subfigure}[t]{0.45\textwidth}
        \centering
        \includegraphics[scale=0.12]{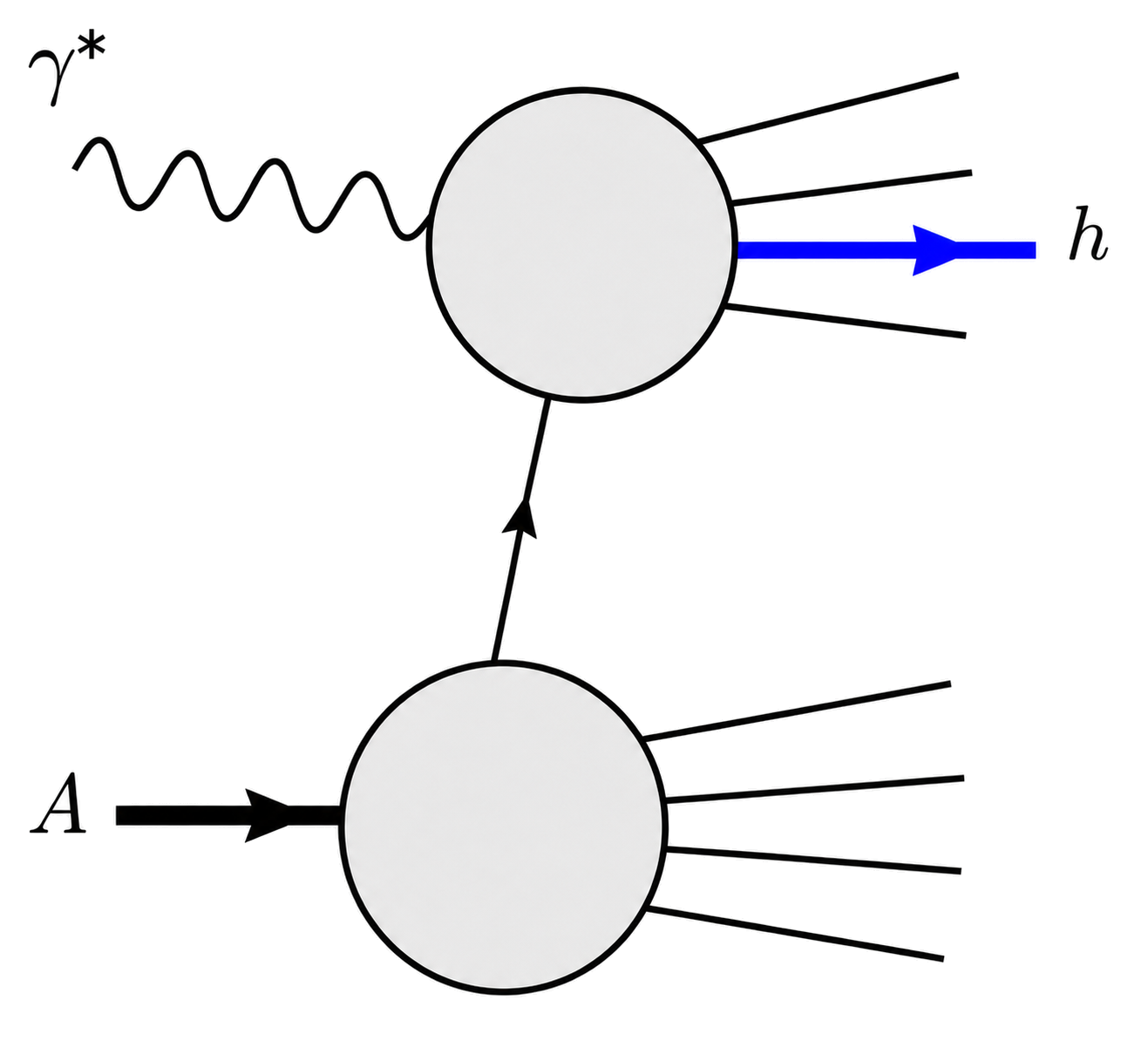}
        \caption{Current fragmentation region}
        \label{fig:SIDIS_CFR}
    \end{subfigure}
     \begin{subfigure}[t]{0.45\textwidth}
        \centering
        \includegraphics[scale=0.12]{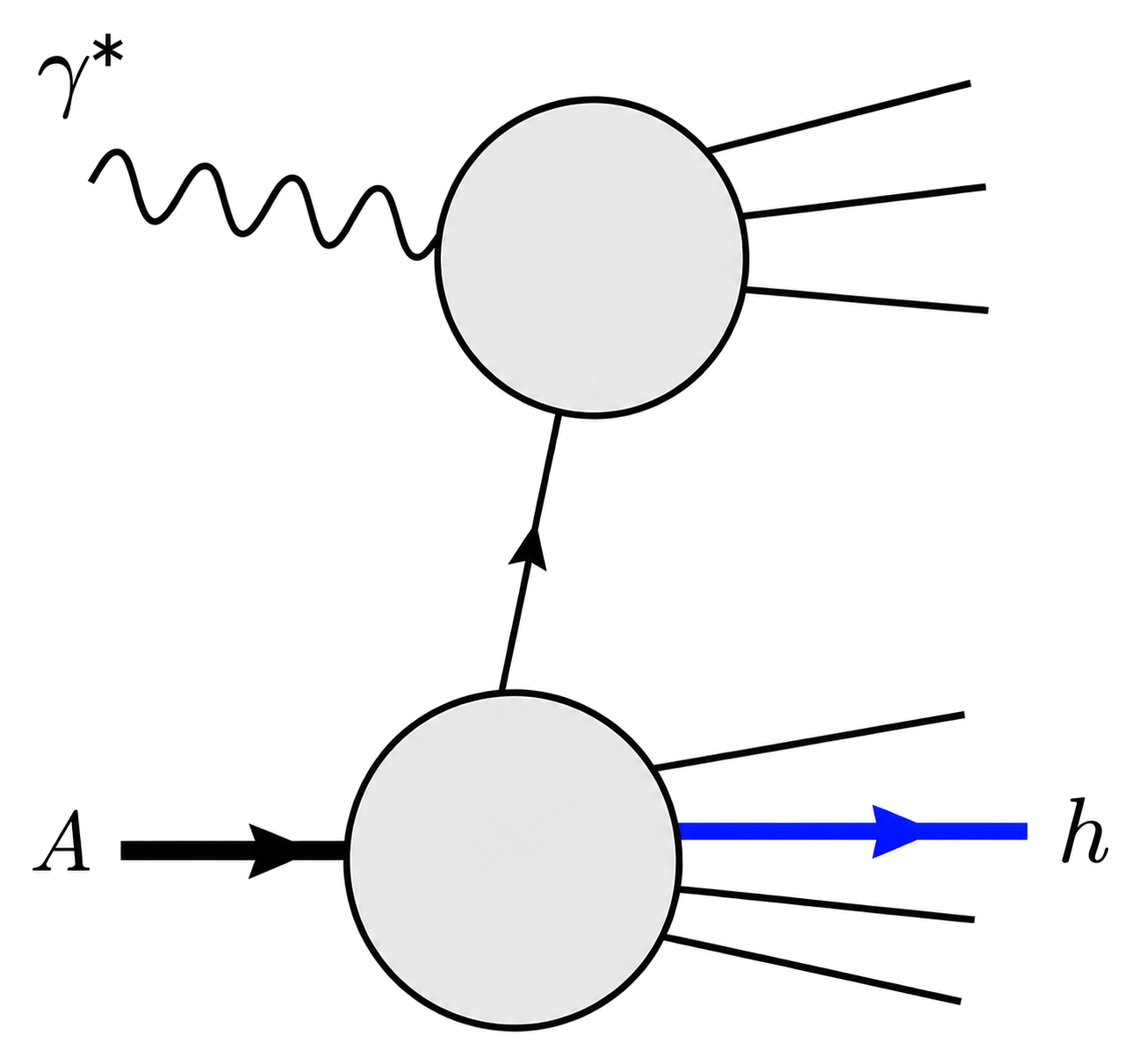}
        \caption{Target fragmentation region}
        \label{fig:SIDIS_TFR}
    \end{subfigure}

  \caption{Schematic illustrations of particle production in the current and target fragmentation regions in DIS.}
    \label{fig:SIDIS_regions}
\end{figure}

By contrast, the target fragmentation region (TFR),  another important production region in DIS~\cite{Boglione:2019nwk}, remains much less explored within the CGC framework. In the TFR, observed particles are produced at relatively small angles with respect to the incoming nucleon beam and  originate predominantly from the target remnant after the removal of an active parton in the hard scattering, as illustrated in Fig.~\ref{fig:SIDIS_regions}.~TFR observables therefore retain nontrivial correlations between the active parton and the observed particles. Such initial--final-state correlations cannot be described by ordinary parton distributions or TMDs, which characterize the initial-state target alone. Instead, they are encoded in \textit{fracture functions}~\cite{Trentadue:1993ka,Berera:1995fj,Grazzini:1997ih,Collins:1997sr,Anselmino:2011ss,Chen:2023wsi,Chen:2024brp} and their inclusive counterparts, \textit{nucleon energy correlators} (NECs)~\cite{Liu:2022wop,Cao:2023oef,Chen:2024bpj}.

Fracture functions provide the conventional framework for nucleon tomography in the TFR~\cite{Trentadue:1993ka,Berera:1995fj,Grazzini:1997ih,Collins:1997sr,Anselmino:2011ss,Chai:2019ykk,Chen:2021vby,Chen:2023wsi,Chen:2024brp,Anselmino:2011bb,Anselmino:2011vkz,Guo:2023uis,Zhao:2024usu,Xi:2025feb,Hatta:2026iry,CLAS:2022sqt}. They describe the conditional distribution of an active parton in the target when a specific hadron $h$ with momentum $P_h$, produced from the target remnant, is observed\footnote{A special class of fracture functions is provided by diffractive parton distributions~\cite{Berera:1995fj,Collins:1997sr}, where a diffractively scattered proton is observed in the final state. Such diffractive fracture functions have recently been studied at small $x$ in~\cite{Iancu:2021rup,Iancu:2022lcw,Hatta:2022lzj,Hatta:2024vzv,Shao:2024nor,Iancu:2025jsu,Shao:2026doo,Caucal:2026xhj}. In this work, however, we focus on the nondiffractive case.
}. The momentum of the detected hadron $P_h$ can correlate with the polarizations of both the active parton and the target, giving rise to a rich set of partonic structures even for collinear fracture functions, where the transverse momentum of the active parton has been integrated out. As a result, collinear fracture functions contain counterparts of familiar TMD distributions,~see the parametrizations in~\cite{Chen:2023wsi,Chen:2024brp}. These include distributions of linearly polarized gluons~\cite{Chen:2023wsi}, encoded through azimuthal correlations with $P_h$, as well as Sivers-type quark distributions~\cite{Anselmino:2011ss,Chen:2021vby} that generate $T$-odd azimuthal asymmetries between the target spin $S$ and $P_h$ in the TFR.
Unlike TMDs, however, fracture functions obey standard DGLAP evolution equations~\cite{Berera:1995fj,Chai:2019ykk} and are thus free from the complications associated with soft gluon radiation in TMD evolution. They thus provide a theoretically clean framework for probing partonic structure, complementary to conventional TMDs. Fracture functions have been widely used in phenomenological analyses of HERA and CLAS data~\cite{Zhao:2024usu,deFlorian:1997wi,deFlorian:1998rj,Shoeibi:2017zha,Shoeibi:2017lrl,Khanpour:2019pzq,CLAS:2022sqt}, as well as to describe a variety of azimuthal correlations in single- and double-hadron-inclusive DIS~\cite{Anselmino:2011ss,Chen:2021vby,Chen:2023wsi,Chen:2024brp,Anselmino:2011bb,Anselmino:2011vkz,CLAS:2022sqt,Guo:2023uis,Hatta:2026iry}.

Inspired by the recent success of energy correlators as probes of QCD dynamics~\cite{Moult:2025nhu}, NECs have emerged as a more inclusive framework for studying nucleon structure through energy-flow measurements in the TFR~\cite{Liu:2023aqb,Mantysaari:2025mht,Liu:2022wop,Cao:2023oef,Cao:2023qat,Liu:2024kqt,Chen:2024bpj,Li:2023gkh,Guo:2024jch,Guo:2024vpe,Huang:2025ljp,Gao:2025cwy,Cao:2026kcg}. Specifically, NECs characterize the correlation between an active parton in the target and the total energy flow in a specified angular direction $(\theta,\phi)$ in the TFR, where $\theta$ and $\phi$ are polar and azimuthal angles, respectively. They offer a simpler experimental means of accessing TFR dynamics while retaining the theoretical advantages of fracture functions. Whereas measurements of fracture functions require the identification and momentum reconstruction of specific hadrons, NECs rely on inclusive calorimetric measurements of the energy deposited within a given angular region. At the same time, an inclusive energy sum rule relates NECs to fracture functions component by component~\cite{Chen:2024bpj}, demonstrating that they encode the same underlying partonic structures and obey similar collinear evolution equations~(see e.g., the evolution studies in  \cite{Cao:2023oef,Gao:2025cwy}). Owing to these features, NECs have been applied to probe spin phenomena~\cite{Chen:2024bpj,Li:2023gkh,Guo:2024jch,Guo:2024vpe,Gao:2025cwy,Huang:2025ljp,Mantysaari:2025mht}, and small-$x$ dynamics~\cite{Liu:2023aqb,Mantysaari:2025mht}, as well as to search for physics beyond the Standard Model~\cite{Huang:2025ljp}.

In DIS, the simplest way to access fracture functions is through single-hadron-inclusive production~(SIDIS) in the TFR~(see e.g.,~\cite{Trentadue:1993ka,Grazzini:1997ih,Collins:1997sr,Berera:1995fj,Anselmino:2011ss,Chen:2023wsi,Chen:2024brp,Hatta:2026iry}). Collinear factorization for this process has been rigorously established in the Bjorken limit~\cite{Collins:1997sr}, and related structure functions have been calculated up to twist-3 level for an unpolarized final-state hadron~\cite{Chen:2023wsi,Chen:2024brp}. Correspondingly, NECs can be measured through the angular distribution of a single energy flow in the TFR in DIS~(see e.g.,~\cite{Liu:2022wop,Cao:2023oef,Chen:2024bpj,Mantysaari:2025mht,Cao:2023qat,Liu:2023aqb}). This distribution is also referred to as the DIS energy pattern~(distribution)~\cite{Meng:1991da,Chen:2024bpj} or the one-point energy correlator in DIS~\cite{Kang:2026hig,Li:2021txc,Kang:2023big}. Using the connection to SIDIS and fracture functions, factorization formulas for various spin- and azimuthal-dependent energy-pattern structure functions have been derived in terms of the corresponding NECs~\cite{Li:2023gkh,Chen:2024bpj}.

With these observables established, calculating fracture functions and NECs within the CGC framework provides a natural avenue for small-$x$ nucleon tomography in the TFR. Recent studies in the quark sector have already demonstrated their sensitivity to gluon saturation~\cite{Liu:2023aqb,Caucal:2025qjg} and $C$-odd odderon effects~\cite{Mantysaari:2025mht}. For an unpolarized target, the unpolarized quark NEC has been expressed in terms of the fundamental dipole $S$-matrix~\cite{Liu:2023aqb}, while the corresponding quark fracture functions have been derived for jet production~\cite{Caucal:2025qjg} and hadron production~\cite{Mantysaari:2025mht}. The analysis has also been extended to a transversely polarized target, establishing a connection between Sivers-type quark NECs and fracture functions and the spin-dependent odderon~\cite{Mantysaari:2025mht}. These leading-twist quark-sector results have so far been obtained at eikonal accuracy.

An initial step toward the gluon sector was taken in Ref.~\cite{Caucal:2025qjg}, which considered fracture functions for jet production and focused on the unpolarized gluon component for an unpolarized target. Rather than starting from the operator definition, Ref.~\cite{Caucal:2025qjg} began with the full CGC description for single-jet production in DIS, took the TFR limit, and matched the result onto the established collinear factorization formula. Surprisingly, the resulting unpolarized gluon jet fracture function was found to exhibit no sensitivity to saturation effects, unlike its quark counterpart. This behavior is not immediately expected, because the unpolarized gluon component is controlled by the adjoint dipole amplitude and might therefore be anticipated to show even stronger saturation effects. It is thus interesting to determine whether this feature persists for hadron fracture functions and their inclusive counterparts, the NECs, particularly through a calculation based directly on their operator definitions.

Analogous to standard gluon TMDs, the leading-twist decompositions of gluon fracture functions~\cite{Chen:2024brp} and gluon NECs~\cite{Liu:2023aqb} each contain eight independent components. Among them, the linearly polarized gluon components are of particular interest because they can generate characteristic azimuthal correlations, providing novel probes of gluon-helicity interference in unpolarized DIS and hard $pp$ collisions. The linearly polarized gluon NEC has been used to describe the long-range near-side ridge at the LHC and has been proposed as a probe of quantum entanglement and possible violations of Bell inequalities~\cite{Li:2023gkh,Guo:2024jch,Guo:2024vpe}. Although linearly polarized gluon TMDs have been studied extensively at small $x$, their fracture-function and NEC counterparts remain unexplored within the CGC framework. The same situation holds for $T$-odd gluon fracture functions and NECs, while their TMD analogues have been shown to be given by the spin-dependent odderon at eikonal accuracy~\cite{Boer:2015pni}.

In this work, we present a systematic calculation of gluonic fracture functions and NECs within the CGC formalism, starting from their operator definitions. We focus on their leading-twist components and evaluate them at eikonal accuracy in the high-energy limit. Using appropriate projection tensors, we first derive gluonic fracture functions and then obtain the corresponding NECs through the inclusive energy sum rule. We find that, among the eight independent gluon components, only the two associated with an unpolarized target survive at this accuracy: the unpolarized and linearly polarized gluon components. Both are determined by the adjoint dipole $S$-matrix. All target-spin-dependent components vanish in the strict eikonal limit. In particular, because the adjoint dipole $S$-matrix is real and $C$-even, no spin-dependent odderon contributes to $T$-odd gluon fracture functions or NECs at this accuracy, in contrast to the quark sector.

We further present a phenomenological study of NEC-based observables at the EIC and demonstrate that the linearly polarized gluon NEC  $h_{1}^{t,g}$ is particularly sensitive to saturation effects.  While the unpolarized gluon NEC  $f_{1}^{g}$ contributes to the azimuthally averaged DIS energy pattern, the linearly polarized gluon NEC generates a characteristic $\cos 2\phi$ modulation in the TFR, where $\phi$ is the azimuthal angle of the energy flow relative to the lepton plane~\cite{Chen:2024bpj}. At leading order in $\alpha_s$ and at small-$x_B$, the resulting asymmetry is governed by the ratio of the {\it linearly polarized gluon} NEC $h_{1}^{t,g}$ to the {\it unpolarized quark} NEC $f_{1}^{q}$ . Within the CGC framework, these two NECs are controlled by the adjoint and fundamental dipole amplitudes, respectively, making their ratio particularly sensitive to the saturation scale. Our numerical results show a substantial suppression of the asymmetry in $e+\mathrm{Au}$ relative to $e+p$ scattering, providing a clear nuclear signature of gluon saturation.

Compared with the dijet $\cos 2\phi$ correlation in the CFR, which is described within TMD factorization, the NEC-induced asymmetry in the TFR offers several theoretical and experimental advantages. Because it is formulated within collinear factorization~\cite{Cao:2023oef,Collins:1997sr}, it is not subject to the Sudakov suppression that typically affects TMD-based dijet observables. It is also free from contributions induced by soft-gluon radiation, which can generate a $\cos 2\phi$ modulation in dijet production and thereby contaminate the intrinsic TMD signal~\cite{Hatta:2020bgy,Hatta:2021jcd,Marquet:2025jdr,Gao:2026azd}. Experimentally, the proposed observable requires only the inclusive measurement of a single energy flow, rather than the reconstruction of a dijet system. Together with the sizable nuclear suppression predicted by our numerical analysis, these features establish the proposed asymmetry as a clean and sensitive probe of the onset of gluon saturation at the future EIC.

The rest of this paper is organized as follows. In Sec.~\ref{sec:op_df}, we introduce the operator definitions of gluon NECs and fracture functions and review their relation through the inclusive energy sum rule. In Sec.~\ref{sec:calculation}, we compute the gluon fracture functions and NECs in the CGC formalism at eikonal accuracy, focusing on the unpolarized and linearly polarized components. In Sec.~\ref{sec:numerics}, we introduce the $\cos 2\phi$ azimuthal asymmetry of the DIS energy pattern as a probe of the linearly polarized gluon NEC and present numerical predictions for EIC kinematics, demonstrating its sensitivity to gluon saturation. Finally, we summarize our findings in Sec.~\ref{sec:conclusion}.

\section{Operator definitions: gluonic NECs and fracture functions}
\label{sec:op_df}

We start by introducing the operator definitions of gluonic NECs and fracture functions, as well as the connection between them. The target nucleon is assumed to move ultrarelativistically in the $+z$ direction, with momentum $P^\mu=(P^+,P^-,\bs 0_\perp)$ such that $P^+\gg P^-$.  We consider the nucleon polarized with the vector $S^\mu$:
\begin{align}
S^\mu = S_L \frac{P^+}{M} \bar n^\mu - S_L \frac{M}{2P^+} n^\mu+ S_\perp^\mu~,
\label{eq:spinvector}
\end{align}
where $S_L$ denotes the nucleon helicity, and $S_\perp^\mu$ represents the nucleon transverse polarization vector. Here, $M$ is the nucleon mass. We employ the light-cone coordinates, in which a four-vector $v^\mu$ is expressed as $v^\mu = (v^+,v^-, \bs v_\perp) $ with $v^{\pm}=(v^0\pm v^3)/{\sqrt{2}}$ and $\bs v_\perp=(v^1, v^2 )$. We also introduce the light-cone vectors $n^\mu = (0,1,0,0)$ and $\bar n^\mu = (1,0,0,0)$, together with the transverse metric tensor defined as $g_\perp^{\mu\nu} = g^{\mu\nu} - \bar n^\mu n^\nu - \bar n^\nu n^\mu$.

The gluonic NECs describe the distribution of a struck gluon with longitudinal momentum fraction $x$ in the target, while simultaneously measuring an energy flow from target remnants in a solid angle~$(\theta,\phi)$, where $\theta$ and $\phi$ are the polar and azimuthal angles, respectively. We focus on the Weizsäcker--Williams-type gluon NECs, which are encoded in the correlation matrix~\cite{Chen:2024bpj,Li:2023gkh}:
\begin{align}
&{\cal M}_{\text{NEC}}^{\alpha \beta} (x,\theta,\phi,S)
\notag \\&~~~~~~=
\frac{1}{x P^{+}} \int \frac{d \eta^-}{2 \pi } e^{-i  x P^{+}\eta^-}\big\langle PS\big|(F^{+\alpha}(\eta) {\cal L}^\dagger(\eta)
)^a{\cal E}(\theta,\phi) \big( {\cal L}(0) F^{+\beta}(0)\big)^a\big|PS\big\rangle\Big \vert_{\eta^+=0,\bs \eta_\perp=\bs 0},
\label{eq:gluonNEEC1}
\end{align}
where the energy-flow measurement is captured by the operator  ${\cal E}( \theta,\phi)$ defined as~\cite{Sveshnikov:1995vi,Bauer:2008dt}
\begin{align}
{\cal E}(\theta,\phi)|X \rangle=\sum_{a \in X} \delta(\theta^2-\theta^2_a)\delta(\phi-\phi_a)\frac{E_a}{E_N}|X \rangle .
\label{eq:Eflow}
\end{align}
Here, $E_a$ denotes the energy of hadron $a$, normalized by the nucleon energy $E_N$, while $F^{\mu\nu}$ denotes the gluon field strength tensor. In Eq.~\eqref{eq:gluonNEEC1}, ${\cal L}(\xi)$ is the light-like gauge link in the adjoint representation, which ensures the gauge invariance of the correlation functions. Physically, this gauge link arises from the final-state or initial-state interactions in a realistic scattering process and extends to either $+\infty$ or $-\infty$ along the light-cone \emph{minus} direction. In this work, we focus on the application in the deep inelastic scattering (DIS) processes, for which the relevant gauge link points toward $+\infty$:
\begin{align}
  {\cal L}_n(\eta^-,\bs \eta_\perp)=[+\infty,\eta^-;\bs \eta_\perp] \, ,
\end{align}
with the Wilson line defined as
\begin{align}
 [\zeta^-,\eta^-;\bs \eta_\perp]=\text{P}~\text{exp}\Big[- i g_s\int^{\zeta^-}_{\eta^-} d z^- A^{+,a}(z^-,\eta_\perp)T^a\Big]~,
 \end{align}
 where $(T^a)_{bc}=if^{bac}$ are the $SU(N_c)$ generators in the adjoint representation.

At leading twist, the NEC correlator, ${\cal M}_{\text{NEC}}^{\alpha \beta}$, from Eq.~\eqref{eq:gluonNEEC1} is parametrized by eight independent gluonic NECs~\cite{Chen:2024bpj}:
\begin{align}
{\cal M}_{\text{NEC}}^{\alpha \beta}= &
- \frac{1}{2}g_\perp^{\alpha \beta} f_{1}^g +\Big( n_{t}^\alpha  n_{t}^\beta +\frac{1}{2}g_\perp^{\alpha \beta} \Big) h_{1}^{t,g}
+ S_L\bigg[i\frac{\varepsilon_\perp^{\alpha \beta}}{2} g_{1L}^g
+ \frac{\tilde  n_{t}^{\{\alpha}  n_{t}^{\beta\}}}{2 } h_{1L}^{t,g}\bigg] \nonumber\\
& + \frac{g_\perp^{\alpha \beta} }{2}
n_{t}\cdot \tilde S_\perp f_{1 T}^{t,g} +n_{t}\cdot S_\perp \bigg[ i\frac{\varepsilon_\perp^{\alpha \beta}}{2} g_{1T}^{t,g}
- \tilde n_{t}^{\{\alpha} n_{t}^{\beta\}} h_{1T}^{tt,g} \bigg]
\notag \\
&
+\frac{\tilde n_{t}^{\{\alpha} S_\perp^{\beta\}}+ \tilde S_\perp^{\{\alpha} n_{t}^{\beta\}}}{4 } h_{1T}^{t,g}~,
\label{eq:gluonNEEC2}
\end{align}
where a unit vector $n_t^\mu$ is introduced to describe the transverse direction of the energy flow:
\begin{align}
n_t^\mu= (0,0,\bs n_t)=(0,0,\cos \phi, \sin \phi)~
\label{eq:nt}.
\end{align}
In the above parametrization, we have used the notation $a^{\{\alpha} b^{\beta\}
 } \equiv a^\alpha b^\beta+a^\beta b^\alpha$ and $\tilde a_\perp^\mu \equiv \varepsilon_\perp^{\mu\nu} a_{\perp\nu}$. The transverse antisymmetric tensor is defined as $\varepsilon_\perp^{\mu\nu} = \varepsilon^{\mu\nu\alpha\beta} \bar n_\alpha n_\beta$ with $\varepsilon^{0123}=1$. Given the decomposition~\eqref{eq:gluonNEEC2}, it is sufficient to calculate the NEC correlator, ${\cal M}_{\text{NEC}}^{\alpha \beta}$, for any transverse indices, $\alpha$ and $\beta$, then apply the appropriate projection to obtain the desired gluonic NEC.

\begin{figure}[t]
  \centering

 \includegraphics[scale=0.17]{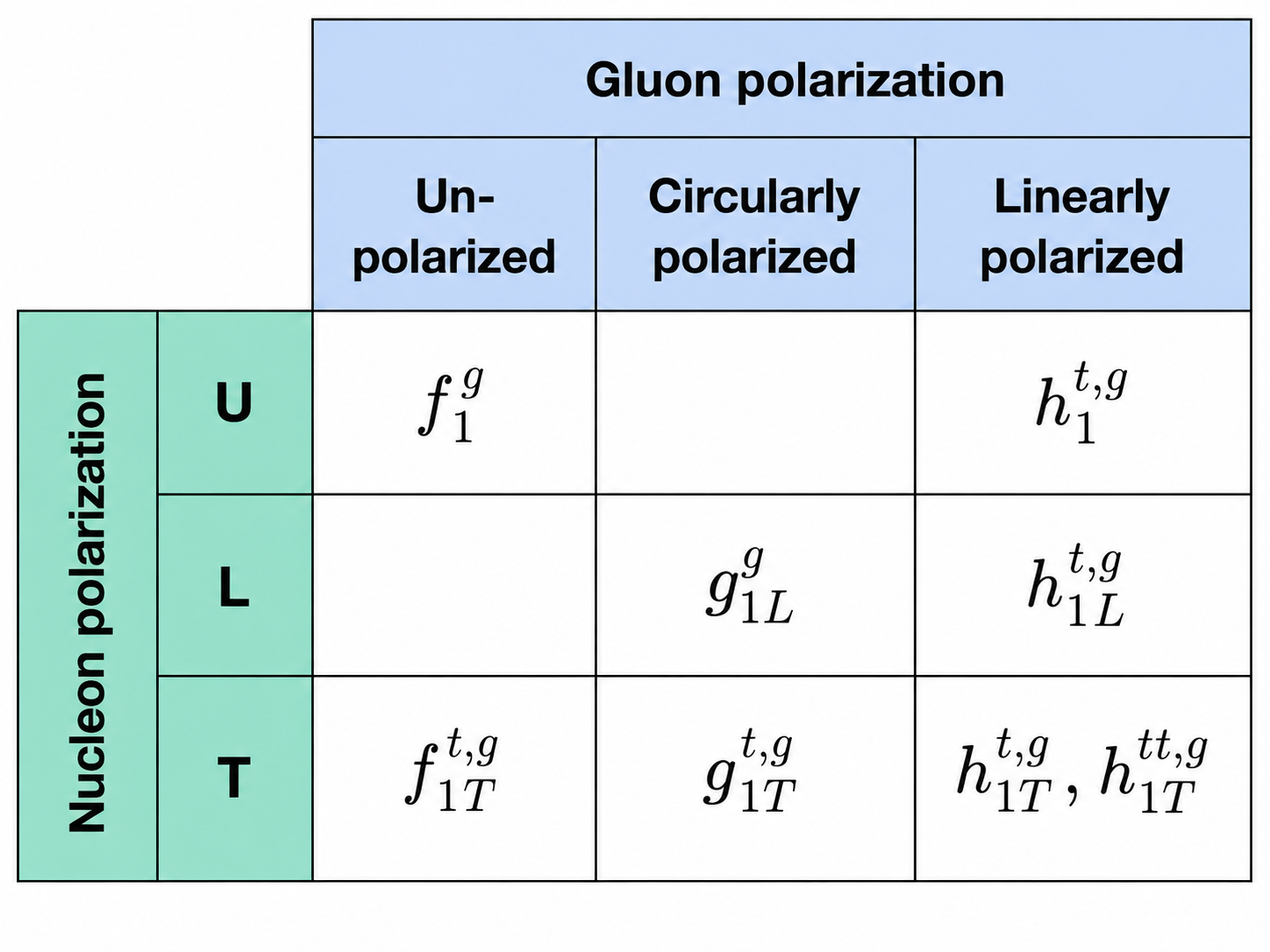}

\caption{Leading-twist gluonic NECs for observing an energy flux from the nucleon target.}
\label{fig:table_gNEC}
\end{figure}

Denoted by functions $f$'s, $g$'s and $h$'s in Eq.~\eqref{eq:gluonNEEC2}, the gluonic NECs are real functions of the gluon momentum fraction $x$ and the polar angle $\theta$ of the measured energy flux. For an unpolarized target, two NECs appear, including $f_1^g$, which describes unpolarized gluons, and $h_1^{t, g}$, which encodes linearly polarized gluons. For a longitudinally polarized nucleon, two additional functions are involved, $g_{1 L}^g$ and $h_{1 L}^{t, g}$ characterizing circularly and linearly polarized gluons, respectively. For a transversely polarized target, the four remaining NECs are relevant,  $f_{1 T}^{t, g}$ describing unpolarized gluons, $g_{1T}^{t,g}$ for circularly polarized gluons, together with $h_{1 T}^{t, g}$ and $h_{1 T}^{t t, g}$ representing two independent structures of linearly polarized gluons correlated with the transverse spin. All  eight gluonic NECs listed above are summarized in Fig.~\ref{fig:table_gNEC} for different spin states of the gluon and the nucleon.

\begin{figure}[t]
  \centering

 \includegraphics[scale=0.17]{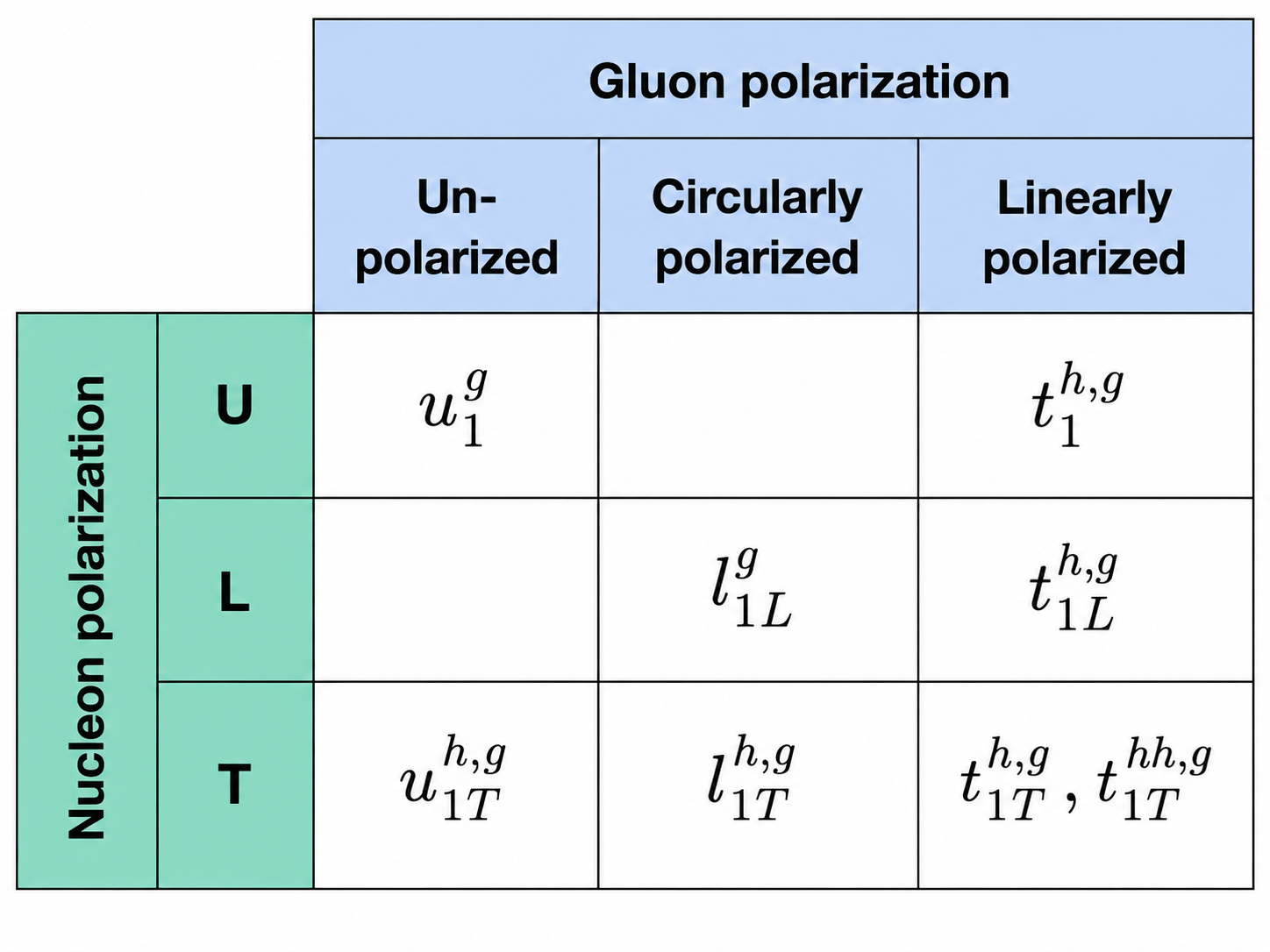}

\caption{Leading-twist gluonic fracture function for observing an unpolarized hadron $h$ from the nucleon target.}
\label{fig:table_gFrF}
\end{figure}

It has been demonstrated in Ref.~\cite{Chen:2024bpj} that gluonic fracture functions can be regarded as the \emph{parent functions} of NECs, as the NEC correlation matrix, ${\cal M}^{\alpha\beta}_{\text{NEC}}(x,\theta,\phi,S)$, can be written in terms of the \emph{gluonic fracture function correlation matrix,} ${\cal M}^{\alpha\beta}_{\text{FrF}}(x,\xi_h,\bs P_{h\perp},S)$, through an inclusive energy sum rule:
\begin{align}
{\cal M}^{\alpha\beta}_{\text{NEC}}(x,\theta,\phi,S)=\sum_h
&\int \frac{d P_{h\perp}^+ d^2\bs P_{h\perp}}{2P_{h}^+(2\pi)^3}\frac{E_h}{E_N}
\delta(\theta^2-\theta_{h}^2)\delta(\phi-\phi_{h}) {\cal M}^{\alpha\beta}_{\text{FrF}}(x,\xi_h,\bs P_{h\perp},S)~,
\label{eq:connection}
\end{align}
where $\sum_h$ represents the summation over all possible species of hadrons. In Eq.~\eqref{eq:connection}, the gluonic fracture function matrix, ${\cal M}^{\alpha\beta}_{\text{FrF}}$, is associated with the measurement of hadron $h$ with momentum $P_h$~\cite{Berera:1995fj,Chen:2024brp}:
\begin{align}\label{eq:gluonMG}
{\cal M}_{\text{FrF}}^{\alpha \beta}(x, \xi_h, \bs P_{h\perp},S) =& \frac{1}{x P^{+}} \int \frac{d \eta^-}{2 \pi} e^{-i
 x P^{+}\eta^-}\sum_X\int\frac{d^3 \bs P_X}{(2\pi)^32 E_X}
\\
&\times \big\langle PS\big|\big(F^{+\alpha}(
\eta ) {\cal L}_n^{\dagger}(
\eta )\big)^a\big|P_{h}, X \big
\rangle \big\langle  P_{h} , X\big|\big({\cal L}_n(0) F^{+\beta}(0)\big)^a\big| PS\big\rangle\Big \vert_{\eta^+=0,\bs \eta_\perp=\bs 0}~.\notag
\end{align}
Since the hadron $h$ is observed in the forward region of the nucleon target, it is convenient to introduce the longitudinal momentum fraction, $\xi_h=P_h^+/P^+$, and write the hadron's momentum as
\begin{align}
P_h^\mu=\Big(\xi_h P^+, \frac{\bs P_{h\perp}^2}{2\xi_h P^+}, \bs P_{h\perp}\Big)~.
\end{align}
At leading twist, the correlator ${\cal M}_{\text{FrF}}^{\alpha \beta}$ contains eight independent gluonic fracture functions similar to the NEC counterpart~\cite{Chen:2024brp}:
\begin{align}
{\cal M}_{\text{FrF}}^{\alpha \beta} = &
- \frac{1}{2}g_\perp^{\alpha \beta} u_{1}^g + \frac{1}{2M^2}\Big( P_{h\perp}^\alpha   P_{h\perp}^\beta + \frac{1}{2}g_\perp^{\alpha \beta} \bs P_{h\perp}^2\Big) t_{1}^{h,g}
+ S_L\bigg[i\frac{\varepsilon_\perp^{\alpha \beta}}{2} l_{1L}^g
+ \frac{\tilde  { P}_{h\perp}^{\{\alpha}  P_{h\perp}^{\beta\}}}{4 M^2} t_{1L}^{h,g}\bigg] \nonumber\\
& + \frac{g_\perp^{\alpha \beta} }{2}
\frac{P_{h\perp}\cdot \tilde S_\perp }{M} u_{1 T}^{h,g} + \frac{P_{h\perp} \cdot S_\perp}{M} \bigg[ i\frac{\varepsilon_\perp^{\alpha \beta}}{2} l_{1T}^{h,g}
- \frac{\tilde P_{h\perp}^{\{\alpha} P_{h\perp}^{\beta\}}}{4M^2} t_{1T}^{hh,g} \bigg]
\notag \\
&
+\frac{\tilde P_{h\perp}^{\{\alpha} S_\perp^{\beta\}}+ \tilde S_\perp^{\{\alpha} P_{h\perp}^{\beta\}}}{8 M} t_{1T}^{h,g}~.
\label{eq:gluonFrF}
\end{align}
Each of the eight fracture functions, denoted by $u$, $l$, and $t$,
 depends on the longitudinal momentum fractions $x$ and $\xi_h$, as well as the squared transverse momentum, $\bs P_{h\perp}^2$. The leading-twist gluonic fracture functions are summarized in Fig.~\ref{fig:table_gFrF} for different spin states of the gluon and the nucleon.

By Eq.~\eqref{eq:connection}, one can establish a one-to-one correspondence between the gluonic NECs and gluonic fracture functions according to the parametrization given in Eq.~\eqref{eq:gluonNEEC2} and Eq.~\eqref{eq:gluonFrF}~\cite{Chen:2024bpj}. As an example, the unpolarized NEC $ f_{1}^g$ can be obtained from the unpolarized fracture function $u_{1}^g$ through:
 \begin{align}
f_{1}^g(x,\theta)=&\sum_{h}\frac{1}{2}
\int_0^{1-x}d \xi_h \xi_h\Big(\frac{\xi_h P^+}{\sqrt{2}}\Big)^2 \frac{1}{(2\pi)^32\xi_h}u_{1}^g
(x,\xi_h,\bs P_{h\perp}^2)
\Big \vert_{\bs P_{h\perp}=\frac{\theta\xi_h P^+}{\sqrt{2}}
 (\cos \phi, \sin \phi)}~.
 \label{eq:sumrule_f1}
\end{align}
Similarly, the linearly polarized gluon NEC, $ h_{1}^{t,g}$, can be calculated from the corresponding fracture function, $t_{1}^{h,g}$, via the relation:
 \begin{align}
   h_{1}^{t,g}(x,\theta)=&\sum_{h}\frac{1}{2}
\int_0^{1-x}d \xi_h \xi_h\Big(\frac{\xi_h P^+}{\sqrt{2}}\Big)^2  \Big( \frac{\theta \xi_h P^+}{\sqrt{2}}\Big)^2\frac{1}{2\xi_h(2\pi)^3}\frac{t_{1}^{h,g}(x,\xi_h,\bs P_{h\perp}^2)}{2M^2}
\Big \vert_{\bs P_{h\perp}=\frac{\theta\xi_h P^+}{\sqrt{2}}
 (\cos \phi, \sin \phi)}~.
  \label{eq:sumrule_h1}
\end{align}
The complete set of correspondences between all the eight pairs of gluonic NECs and fracture functions can be found in Ref.~\cite{Chen:2024bpj}.

\section{Gluonic NECs and fracture functions at small-$x$  }
\label{sec:calculation}

In this section, we compute the gluonic NECs and fracture functions at small $x$ within the shockwave formalism of the CGC effective theory~\cite{Iancu:2003xm,Gelis:2010nm,Kovchegov:2012mbw,Morreale:2021pnn}, where the fast-moving target acts as a dense source of small-$x$ gluons. These gluons are characterized by the semi-hard saturation scale $Q_s$ and typically carry transverse momentum $k_{g\perp}\lesssim Q_s$, with $Q_s \gg \Lambda_{\text{QCD}}$. Consequently, the correlations between the initial parton and the energy flow encoded in the NECs become perturbatively calculable. Such small-$x$ effects can be probed by measuring the energy flow in the angular region, $\theta Q \lesssim Q_s$. In this work, we restrict the calculation to leading order~(LO) in $\alpha_s$  and to the leading power within the high-energy limit ($P^+\propto \sqrt{s}\rightarrow +\infty$) with $s$ as the squared center-of-mass energy of the DIS process. The latter corresponds to the eikonal limit in the CGC language.

\subsection{Evaluations in the shockwave formalism}

We begin by evaluating the gluonic fracture function associated with the production of a real gluon carrying momentum $p_g$. For practical computations in the CGC framework, it is convenient to use translational invariance to recast the correlator in Eq.~(\ref{eq:gluonMG}) into the form: \begin{align}
{\cal M}_{G,\text{FrF}}^{\alpha \beta} (x,\xi,\bs p_{g\perp})=&\frac{2P^+}{\ell^+(2\pi)^3\langle P |P\rangle}\sum_X\int d^2 \bs \ell _\perp  \int d^3 \eta \, d^3\zeta \, e^{i (\eta-\zeta) \cdot \ell}
 \label{eq:Gfracsmallx} \\
&\times \langle P|\big(F^{+\alpha}(\zeta) {\cal L}_n^{\dagger}(\zeta)\big)^a |p_{g},X\rangle \langle
 X,p_{g}| \big({\cal L}_n(\eta) F^{+\beta}(\eta)\big)^a|P\rangle\Big\vert_{\eta^+,\zeta^+=0}~
 \notag \\
 =& \frac{2P^+}{\ell^+(2\pi)^3\langle P |P\rangle}\int d^2 \bs \ell _\perp \sum_X \sum_{\lambda}\left[{\cal A}^{ac,\alpha}(x,\xi,\bs \ell_\perp,\lambda)\right]^*{\cal A}^{ac,\beta}(x,\xi,\bs \ell_\perp,\lambda)~,
\notag
\end{align}
where $\xi=p_{g}^+/P^+$, $\ell^\mu=(x P^+,0,\bs {\ell }_\perp)$, and $\langle P |P\rangle=2P^+(2\pi)^3\lim_{\Delta\rightarrow0}\delta(\Delta^+)\delta^{(2)}(\Delta_\perp)$. Here, integration measure $d^3\eta$ extends over the full ranges of $\eta^-$ and $\bs \eta_\perp$. In the final step of Eq.~\eqref{eq:Gfracsmallx}, we also introduced the fracture–function amplitude,
\begin{align}
{\cal A}^{ac,\beta}(x,\xi,\bs \ell_\perp,\lambda)=\int d^3 \eta \;  e^{i \eta \cdot \ell}
 \langle
 X,\{p_{g},\lambda,c\}| \big({\cal L}_n(\eta) F^{+\beta}(\eta)\big)^a|P\rangle\Big\vert_{\eta^+=0}~,
 \label{eq:amp_def}
\end{align}
which describes the emission of a real gluon with color $c$ and polarization $\lambda$.

\subsubsection{Calculations at the amplitude level}

In the CGC framework, the fast-moving target acts as a dense source of small-$x$ gluons, and its interaction with propagating partons is encoded in an instantaneous shockwave. Here, we assume that the target moves along the $\hat z$-axis with a large plus momentum component $P^+$. Throughout our calculation, we work in the light-cone gauge $A^- = 0$. In this gauge, for a fast-moving gluon projectile carrying a large minus momentum $p^-_g$, its  interaction with the shockwave at the transverse position $\bs z_\perp$ is represented by a light-like Wilson line in the adjoint representation~\cite{Iancu:2003xm},
\begin{align}
U(\bs z_\perp)
=
[+\infty,-\infty;\bs z_\perp]
=
\mathcal{P}\exp\Big[-ig_s\int_{-\infty}^{+\infty}dz^-\, A^{+,a}(z^-,\bs z_\perp)\,T^a\Big]\,.
\label{eq:Uline}
\end{align}
Here, $A^{+,a}$ denotes the background CGC gluon field, which is nonvanishing only within a narrow interval around $z^-=0$. For $p^-_g>0$, the corresponding CGC vertex is given as~\cite{Iancu:2003xm}
\begin{align}\label{CGC_gl_vertex}
  -g_{\mu\nu} 2 p_g^-
 \int d^2\bs z_\perp  e^{-i \bs z_\perp\cdot \bs k_{g\perp}}\big[ U(\bs{z}_{\perp}) \big]_{ab}~,
\end{align}
where $\{\mu,a\}$ and $\{\nu,b\}$ are the Lorentz and color indices  of the fast-moving gluons before and after the shockwave interactions, respectively. Here, $\bs k_{g\perp}$ denotes the net transverse momentum transferred from the target small-$x$ gluon field. In the case where $p_g^- < 0$, the gluon projectile corresponds instead to the Hermitian conjugate of $U$, and one must replace $U(\bs z_\perp)$ in Eq.~\eqref{CGC_gl_vertex} by $-U^\dagger(\bs z_\perp)$.

As usual, an internal gluon line of momentum $p_g$ receives the factor of gluon propagator, $i\Pi_{\mu\nu}(p_g)$, where
\begin{align}
    \Pi^{\mu\nu}(p_g) = -g^{\mu\nu} + \frac{\bar{n}^{\mu}p^{\nu}_g + \bar{n}^{\nu}p^{\mu}_g}{\bar{n}\cdot p_g}
\end{align}
and as a reminder $\bar{n} = (1,0,0_{\perp})$. Furthermore, an outgoing on-shell gluon with momentum $p_g$ and polarization $\lambda$ yields the polarization four-vector,
\begin{align}
\varepsilon^{\mu*}_{\lambda}(p_g) = \left(\frac{\varepsilon^*_{\lambda\perp}\cdot\bs{p}_{g\perp}}{\bar{n}\cdot p_g},0,\bs \varepsilon^*_{\lambda\perp}\right),
\end{align}
and the physical polarization sum satisfies
\begin{align}
\sum_{\lambda=\pm1} \epsilon^{\mu*}_\lambda(p_g) \epsilon^{\nu}_\lambda (p_g)=\Pi^{\mu\nu}(p_g)~.
\end{align}

In the amplitude ${\cal A}^{ac,\beta}$ given in Eq.~\eqref{eq:amp_def}, the observed real gluon is generated through the interaction of the operator ${\cal L}_n F^{+\beta}$ with this shockwave background. The gauge link ${\cal L}_n$ has a clear physical interpretation: it describes the trajectory of a fast final-state gluon propagating opposite to the target. As it traverses the small-$x$ gluon field, it undergoes multiple scatterings and therefore already incorporates part of the shockwave interaction with the projectile. In particular, within the limit considered in our calculation, we write
\begin{align}
    \mathcal{L}_n(\eta) &= \mathbb{I}\,\theta(\eta^-) + U(\bs\eta_{\perp})\,\theta(-\eta^-)\,,
\end{align}
such that the gauge link interacts with the shockwave at the location, $\eta^-=0^-$.

\begin{figure}[h]
  \centering

\includegraphics[scale=0.6]{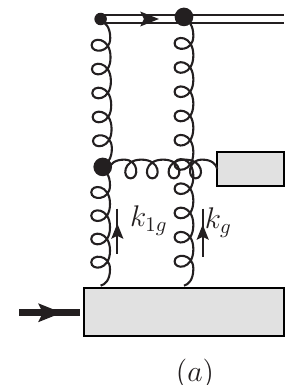}
      \hspace*{1cm}
\includegraphics[scale=0.6]{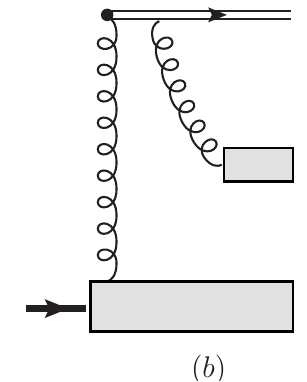}

\caption{CGC amplitudes for producing a gluon jet in gluonic NECs or fracture functions, in the cases where (a) the real gluon is emitted before the shockwave interaction, and (b) the real gluon is emitted after the shockwave interaction. The double line represents the gauge link which goes to $+\infty$ in the DIS process.}
\label{fig:gNEC_amp}
\end{figure}

At LO in perturbation theory, two diagrams shown in Fig.~\ref{fig:gNEC_amp} contribute to ${\cal A}^{ac,\beta}$.
Fig.~\ref{fig:gNEC_amp}(a) corresponds to the case where the real gluon is emitted before the shockwave interaction with the target field. The second diagram, Fig.~\ref{fig:gNEC_amp}(b), represents the configuration where the real gluon is emitted after passing through the shockwave. In each diagram, the smaller rectangle represents the emitted real gluon, while the larger rectangle represents the target nucleon. In particular, the vertical gluon lines represent the shockwave interactions, which at the eikonal level include exchanges of any number of gluons between the projectile and the target. The calculations of these NEC diagrams
 follow similar techniques as used for TMDs, as detailed in~\cite{Xiao:2017yya, Kovchegov:2015pbl,Kovchegov:2018znm,Cougoulic:2022gbk,Adamiak:2024khm}.

For the contributions from the real gluon emitted before the shockwave in Fig.~\ref{fig:gNEC_amp}(a), the fracture–function amplitude ${\cal A}^{ac,\beta }$ is expressed as
 \begin{align}\label{A_2a_1}
{\cal A}^{ac,\beta }\Big\vert_{\text{Fig}.~\ref{fig:gNEC_amp}(a)}
=
&\int\frac{ d k_g^+ d^2\bs k_{g\perp}}{(2\pi)^3 } \int \frac{d k_{1g}^+ d^2\bs k_{1g\perp}}{(2\pi)^3} (2\pi)^3 \delta^{(3)}[\ell-(k_g+k_{1g}-p_g)] \frac{i}{-k_g^++i\epsilon}
\notag \\
& \times (-i)\big[ (k_g+k_{1g}-p_g)^+g^{\beta \nu}_\perp-(k_{1g}-p_g)^\beta_\perp n^\nu\big] \frac{i\Pi_{\nu \sigma}(k_{1g}-p_g)2 p_g^- \epsilon^{\sigma*}_\lambda  (p_g) }{( k_{1g}- p_g)^2+i\epsilon}
 \notag \\
&\times \int d^2 z_\perp d^2y_\perp  e^{-i (\bs z_\perp\cdot \bs k_{g\perp}+\bs y_\perp\cdot \bs k_{1g\perp})}
 (-1)\langle X|\big[ U(\bs z_\perp)  U^\dagger(\bs y_\perp) -1\big]_{ac}|P\rangle~,
\end{align}
where the delta function enforces the conservation of the $+$ and $\perp$ momenta and can be used to eliminate the integration over $k_{1g}$. The internal gluon propagator together with the vertex factor are expressed in the second line of Eq.~\eqref{A_2a_1}, while the last line represents the eikonal interaction within the target shockwave. As the next step, we evaluate the integral over $k_g^{+}$ in Eq.~\eqref{A_2a_1} using the residue theorem, which picks up the pole at $k_g^{+}=i\epsilon$. As a result, we obtain
 \begin{align}
{\cal A}^{ac,\beta }\Big\vert_{\text{Fig}.~\ref{fig:gNEC_amp}(a)}
=
&\int\frac{  d^2\bs k_{g\perp}}{(2\pi)^2 } (-i)  \big[ (k_{1g}-p_g)^+g^{\beta \nu}_\perp-(k_{1g}-p_g)^\beta_\perp n^\nu\big] \frac{i\Pi_{\nu \sigma}(k_{1g}-p_g) 2 p_g^- \epsilon^{\sigma*}_\lambda  (p_g) }{-2\ell^+p_g^--(\bs \ell- \bs k_{g\perp})_\perp^2}
 \notag \\
&\times \int d^2 z_\perp d^2y_\perp  e^{-i (\bs z_\perp\cdot \bs k_{g\perp}+\bs y_\perp\cdot \bs k_{1g\perp})}
 (-1)\langle X|\big[U(\bs z_\perp) U^\dagger(\bs y_\perp) -1\big]_{ac}|P\rangle~,
 \label{eq:amp(a)}
\end{align}
where $k_{1g}^+=\ell^++p_g^+$ and $\bs k_{1g\perp}=\bs \ell_\perp-\bs k_{g\perp}+\bs p_{g\perp}$.

We now move to the contributions from the real gluon emission after the shockwave interaction shown in Fig.~\ref{fig:gNEC_amp}(b), in which the emitted gluon attaches to the gauge link ${\cal L}_n$.  In this case, the resulting amplitude takes the form:
\begin{align}
{\cal A}^{ac,\beta}\Big\vert_{\text{Fig}.\ref{fig:gNEC_amp}(b)}
 =\int d^2 \bs\eta_\perp  e^{-i \bs  \eta_\perp \cdot(\bs \ell+\bs p_{g\perp} )_\perp}\frac{2\bs{\epsilon}_{\perp}^{*\lambda} \cdot \bs{p}_{g\perp}}{\bs p_{g\perp}^2}
 \langle
 X|U^{cd}(\bs \eta) (i\partial_{
\bs \eta_\perp}^\beta U^\dagger)^{da}(\bs \eta)|P\rangle~.
\label{eq:amp(b)}
\end{align}
To derive this, we start from the original fracture–function amplitude in Eq.~\eqref{eq:amp_def}. In order to contract the external gluon state $ \langle\{p_{g},\lambda,c\}|$ with the gauge link, we need to employ the identity,
 \begin{align}
  {\cal L}_n(\eta^-,\bs \eta_\perp)=1-i g_s \int^{+\infty}_{ \eta^-}  d \xi^- A^+(\xi^-,\bs \eta_\perp)[\xi^-, \eta^- ;\bs \eta_\perp]~,
\end{align}
to make the $A^+$-field explicit and have the creation operator to contract with the one-gluon bra state. After this contraction, we obtain
\begin{align}
{\cal A}^{ac,\beta}\Big\vert_{\text{Fig}.\ref{fig:gNEC_amp}(b)}
=&\int d^2\bs \eta_\perp  e^{i \ell^+\eta^--i\bs \eta \cdot \bs \ell_\perp} \int^{+\infty}_{-\infty} d\eta^- \int ^{+\infty}_{\eta^-}d \xi^-   e^{ip^+_g\xi^--i\bs p_{g\perp} \cdot \bs \eta_\perp} n\cdot\epsilon^*_\lambda(p_g)
\notag \\
&\times
 \langle
 X| (-i g_s)(T^c)^{ae} [\xi^-, \eta^- ;\bs \eta_\perp]^{eb} F^{+\beta,b}(\eta)|P\rangle~.
\end{align}
We next apply the approximation $e^{i \ell^+\eta^-}=e^{ixP^+\eta^-}\approx 1$, whose corrections are beyond the eikonal level~\cite{Cougoulic:2022gbk,Kovchegov:2025gcg}, and interchange the order of integration using $    \int^{+\infty}_{-\infty} d\eta^- \int ^{+\infty}_{\eta^-}d \xi^-
  =  \int^{+\infty}_{-\infty} d \xi^- \int ^{\xi^-}_{-\infty}d\eta^-$.
Then, the $\eta^-$-integration yields
\begin{align}
  &(-i g_s)\int ^{\xi^-}_{-\infty}d\eta^-  (T^c)^{ae} [\xi^-, \eta^- ;\bs \eta_\perp]^{eb} F^{+\beta,b}(\eta)
=\big(\partial_{\perp}^\beta [\xi^-,-\infty;\bs \eta_\perp]^{ab}\big)[-\infty ,\xi^-;\bs \eta_\perp]^{bc}~.
\end{align}
This leaves us with
\begin{align}
{\cal A}^{ac,\beta}\Big\vert_{\text{Fig}.\ref{fig:gNEC_amp}(b)}
=&\int d^2\bs \eta_\perp  e^{-i\bs \eta \cdot \bs \ell_\perp} \int^{+\infty}_{-\infty} d \xi^-   e^{ip^+_g\xi^--i\bs p_{g\perp} \cdot \bs \eta_\perp} n\cdot\epsilon^*_\lambda(p_g)
\notag \\
&\times \langle
 X| \Big(\partial_{\perp}^\beta [\xi^-,-\infty;\bs \eta_\perp]^{ab}\Big)[-\infty ,\xi^-;\bs \eta_\perp]^{bc}|P\rangle~.
\end{align}
To proceed with the $\xi^-$-integration, we notice that $n\cdot\epsilon^*(p_g)=2p^+_g\frac{\bs{\epsilon}_{\perp}^{*\lambda} \cdot \bs{p}_{g\perp}}{\bs p_{g\perp}^2}$ and $p_g^+ e^{ip^+_g\xi^-}=-i\frac{\partial}{\partial\xi^-}e^{ip^+_g\xi^-}$. Together with integration by parts, we obtain
 \begin{align}
{\cal A}^{ac,\beta}\Big\vert_{\text{Fig}.\ref{fig:gNEC_amp}(b)}
=&\int d^2\bs \eta_\perp  e^{-i\bs \eta \cdot \bs \ell_\perp} \int^{+\infty}_{-\infty} d \xi^-   e^{ip^+_g\xi^--i\bs p_{g\perp} \cdot \bs \eta_\perp}\frac{\bs{\epsilon}_{\perp}^{*\lambda} \cdot \bs{p}_{g\perp}}{\bs p_{g\perp}^2}
\notag \\
&\times2 i\partial_{\xi^-}
 \langle
 X| \Big(\partial_{\perp}^\beta [\xi^-,-\infty;\bs \eta_\perp]^{ab}\Big)[-\infty ,\xi^-;\bs \eta_\perp]^{bc}|P\rangle~.
\end{align}
Finally, we once again make the eikonal approximation, $e^{i p_g^+ \xi^-} \approx 1$, which renders the $\xi^{-}$-integration trivial. Then we arrive at the compact expression given in Eq.~\eqref{eq:amp(b)}, which holds at the eikonal level.

\subsubsection{Squaring the CGC amplitudes}

\begin{figure}[htbp]
  \centering

  \begin{subfigure}[b]{0.29\textwidth}
\includegraphics[scale=0.50]{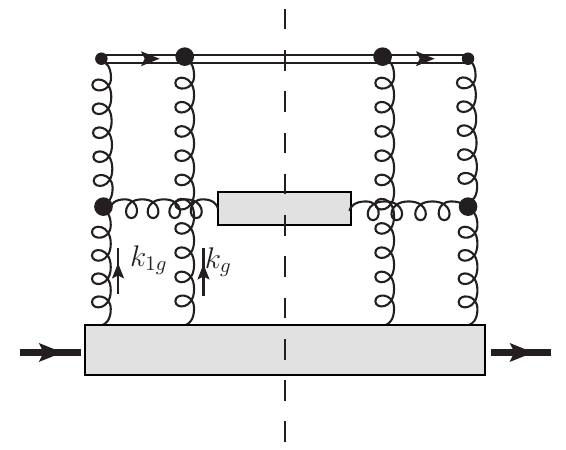}
\caption{}
    \label{fig:ampSquare_A}
  \end{subfigure}
   \begin{subfigure}[b]{0.29\textwidth}
\includegraphics[scale=0.50]{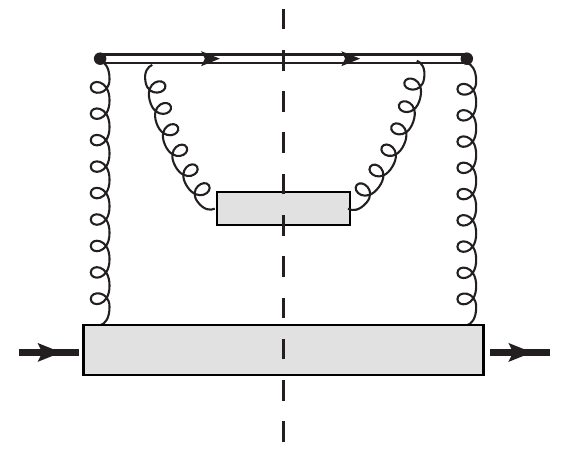}
\caption{}
    \label{fig:ampSquare_B}
  \end{subfigure}
    \begin{subfigure}[b]{0.29\textwidth}
\includegraphics[scale=0.50]{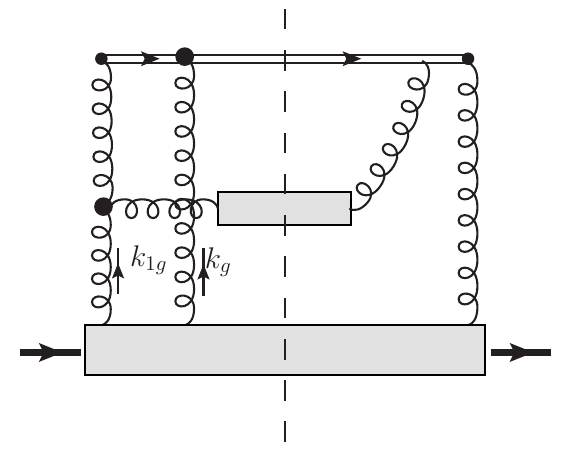}
\caption{}
    \label{fig:ampSquare_C}
  \end{subfigure}
\caption{Contributions to the gluonic NECs and fracture functions at small-$x$. Mirror diagrams are implied.}
\label{fig:gluonNEEC}
\end{figure}

In light of Eq.~\eqref{eq:Gfracsmallx}, the gluonic fracture functions and hence the gluonic NECs can be calculated by \emph{squaring} the fracture-function amplitude, ${\cal A}^{ac,\beta}$. The latter contains the two diagrams shown in Figs.~\ref{fig:gNEC_amp}(a) and \ref{fig:gNEC_amp}(b), whose amplitudes are given via CGC formalism in Eqs.~\eqref{eq:amp(a)} and \eqref{eq:amp(b)}, respectively. The fracture functions, ${\cal M}_{\text{FrF}}^{\alpha \beta}$, and hence the NECs correspond to the three diagrams shown in Fig.~\ref{fig:gluonNEEC}, which correspond respectively to the square of Fig.~\ref{fig:gNEC_amp}(a), the square of Fig.~\ref{fig:gNEC_amp}(b) and the interference terms.

We find that the results can be expressed in terms of the dipole gluon distribution in the adjoint representation:
  \begin{align}
   {\cal F}_{x_g}(\bs k_{g\perp})=\int \frac{d^2\bs b_{\perp} d^2 \bs r_{\perp}}{(2 \pi)^2} e^{-i \bs k_{g\perp} \cdot\bs  r_{\perp}} {\cal S}_{x_g}(\bs r_\perp, \bs b_\perp)
   ,
 \label{eq:gluondipole}
 \end{align}
  which is defined as the Fourier transform of the gluon dipole $S$-matrix:
\begin{align}
 {\cal S}_{x_g}(\bs r_\perp, \bs b_\perp) =\frac{1}{N_c^2-1}\Big \langle \text{Tr}
\Big[U(\bs{b}_{\perp}+\frac{\bs{r}_{\perp}}{2}) U^{\dagger}(\bs{b}_{\perp}-\frac{\bs{r}_{\perp}}{2})\Big] \Big\rangle_{x_g}~.
\label{eq:gluondipole_coordinate}
\end{align}
Here, $U(\bs z_\perp)$ is the adjoint Wilson line defined in Eq.~(\ref{eq:Uline}), and $\langle \cdots \rangle_{x_g}$ denotes the CGC average of the target color sources.

The contribution from the diagram in Fig.~\ref{fig:gluonNEEC}(A) is obtained by squaring the amplitude in Eq.~\eqref{eq:amp(a)}, which eventually simplifies to\begin{align}
x{\cal M}_{\text{FrF}}^{\alpha \beta}  (x,\xi,\bs p_{g\perp})\Big\vert_{(A)}
=&-\frac{4(N_c^2-1)}{\pi}
\int d^2\bs k_{g\perp}{\cal F}_{x_g}(\bs k_{g\perp})
     \bigg[
\frac{{\cal N}^{\alpha \gamma }(\bs p_{g\perp}-\bs k_{g\perp})}{\epsilon_f^2+(\bs p_{g\perp}-\bs k_{g\perp})^2} -\frac{{\cal N}^{\alpha \gamma }(\bs p_{g\perp})}{\epsilon_f^2+ \bs p_{g\perp}^2}  \bigg]
\notag \\
&\times
\bigg[
\frac{{\cal N}_\gamma^{~\beta}(\bs p_{g\perp}-\bs k_{g\perp})}{\epsilon_f^2+(\bs p_{g\perp}-\bs k_{g\perp})^2} -\frac{{\cal N}_\gamma^{~\beta}(\bs p_{g\perp})}{\epsilon_f^2+ \bs p_{g\perp}^2}  \bigg] ~,
 \label{eq:diagramA}
\end{align}
where $\epsilon_f^2\equiv(x/\xi) \bs p^2_{g \perp}$, and ${\cal N}^{\alpha \gamma }(\bs k_\perp)= \bs k_\perp^\alpha \bs k_\perp^\gamma-\frac{g^{\alpha \gamma}_\perp }{2}\epsilon_f^2$.

Next, the contribution from Fig.~\ref{fig:gluonNEEC}(B), obtained by squaring the amplitude in Eq.~\eqref{eq:amp(b)}, is of the form:
\begin{align}
  x{\cal M}_{\text{FrF}}^{\alpha \beta}(x,\xi,\bs p_{g\perp})  \Big \vert_{(B)}
=&\frac{4(N_c^2-1)}{\pi} \int d^2 \bs k _{g\perp}\frac{ \bs k_{g\perp}^\alpha \bs k_{g\perp}^\beta }{\bs p_{g\perp}^2}  {\cal F}_{x_g}(\bs k_{g\perp})~.
  \label{eq:diagramB}
\end{align}
Finally, the diagram in Fig.~\ref{fig:gluonNEEC}(C), together with its mirror diagram, represents the interference between the amplitudes of Eqs.~\eqref{eq:amp(a)} and \eqref{eq:amp(b)}. This is given by
\begin{align}
x{\cal M}^{\alpha \beta}_{\text{FrF}}&(x,\xi,\bs p_{g\perp}) \Big \vert_{\text{(C)}}
\notag \\
=&-\frac{4(N_c^2-1)}{\pi} \frac{1}{\bs p_{g\perp}^2} \int d^2\bs k_{g\perp} \bigg\{\bs k_{g\perp }^\alpha(\bs k_{g\perp}-\bs p_{g})^\beta_\perp\frac{1}{2}\Big[
\epsilon_f^2+(\bs k_{1g}- \bs p_{g})^2_\perp+\bs p_{g\perp}^2-\bs k_{g\perp}^2\Big]-\bs k_{g\perp }^\alpha\bs k_{g\perp}^\beta \frac{\epsilon_f^2}{2}\bigg\}
\notag \\
&\times\frac{1}{\epsilon_f^2+(\bs k_{g}- \bs p_{g})_\perp^2}
 {\cal F}_{x_g}(\bs k_{g\perp})+\{\alpha\leftrightarrow\beta\}~.
  \label{eq:diagramC}
 \end{align}
Then, the correlation matrix, $x{\cal M}^{\alpha \beta}_{\text{FrF}}$, is simply the sum of the contributions in Eqs.~\eqref{eq:diagramA},~\eqref{eq:diagramB} and ~\eqref{eq:diagramC}.

 \subsection{Results and Discussions}
 \label{sec:CGC_results}

As expressed at the end of the previous Section, Eqs.~\eqref{eq:diagramA},~\eqref{eq:diagramB} and ~\eqref{eq:diagramC} sum to the final expression for the correlation matrix, $x{\cal M}^{\alpha \beta}_{\text{FrF}}$, in the CGC formalism. With this result in hand, one can now project out the individual components of the gluonic fracture functions and subsequently derive the corresponding NECs. Among the eight components presented in Eqs.~\eqref{eq:gluonNEEC2} and \eqref{eq:gluonFrF}, only the structures associated with an unpolarized target survive in the strict eikonal limit. All contributions sensitive to the target spin are beyond eikonal.  This follows from the fact that the eikonal gluon-dipole $S$-matrix is nonvanishing only for an unpolarized target.

First,
eikonal interactions preserve the helicity of the target, rendering the associated  production rates insensitive to the target's longitudinal spin~\cite{Kovchegov:2017lsr,Cougoulic:2022gbk,Kovchegov:2021iyc}. On the other hand, unlike its quark-dipole counterpart, the eikonal gluon-dipole $S$-matrix is strictly real. As a consequence, it cannot contain an odderon contribution, which is purely imaginary and constitutes the only source of target transverse-spin effects at eikonal accuracy. This is in contrast to the case of quark NECs studied previously in Ref.~\cite{Mantysaari:2025mht}, where an unpolarized quark NEC in a transversely polarized target can already arise at the eikonal level due to the spin-dependent odderon.

Therefore, only the unpolarized and linearly polarized gluon fracture functions and NECs receive eikonal contributions. We summarize their results below.

\subsubsection{Unpolarized gluon fracture function and NEC}

  In this section, we extract the unpolarized gluon fracture function, $u_{1}^g$, from our results in Eqs.~\eqref{eq:diagramA} to \eqref{eq:diagramC}. Given the decomposition~\eqref{eq:gluonFrF} of the fracture function, ${\cal M}_{G,\text{FrF}}^{\alpha \beta}$,
 we see that $u_{1}^g$ is projected via
  \begin{align}
     u_{1}^g=&-g^{\alpha\beta}_\perp{\cal M}_{\text{FrF},\alpha \beta}~. \label{eq:pro1}
       \end{align}
  All three diagrams in Fig.~\ref{fig:gluonNEEC} contribute under this projection. Applying Eq.~\eqref{eq:pro1} to Eqs.~\eqref{eq:diagramA}-\eqref{eq:diagramC}, we respectively get
        \begin{align}
     xu_{1}^g\bigg\vert_{(A)}
=&\frac{4(N_c^2-1)}{\pi}
\int d^2\bs k_{g\perp} {\cal F}_{x_g}(\bs k_{g\perp})\Big[   {\cal T}_g(\bs p_{g\perp}-\bs k_{g\perp}, \bs p_{g\perp}-\bs k_{g\perp})-2{\cal T}_g(\bs p_{g\perp}-\bs k_{g\perp}, \bs p_{g\perp})
\notag \\
&\hspace{5cm}  +{\cal T}_g(\bs p_{g\perp}, \bs p_{g\perp})
 \Big]  ,
 \\
     xu_{1}^g\bigg\vert_{(B)}
=&\frac{4(N_c^2-1)}{\pi} \frac{1  }{\bs p_{g\perp}^2}
\int d^2\bs k_{g\perp}  \bs k_{g\perp }^2  {\cal F}_{x_g}(\bs k_{g\perp})~,
 \\
 xu_{1}^g \Big \vert_{(C)}
=&-\frac{4(N_c^2-1)}{\pi} \frac{1}{\bs p_{g\perp}^2} \int d^2\bs k_{g\perp} \left[\bs k_{g\perp }^2
+\frac{\bs k_{g\perp }\cdot(\bs k_{g\perp}-\bs p_{g})_\perp(\bs p_{g\perp}^2-\bs k_{g\perp}^2)-\bs k_{g\perp}^2 \epsilon_f^2}{\epsilon_f^2+(\bs k_{g\perp}-\bs p_{g\perp})^2}\right] {\cal F}_{x_g}(\bs k_{g\perp})
~,
\end{align}
where $\epsilon_f^2=x\bs p_{g\perp}^2/\xi$, and the coefficient function ${\cal T}_g$ is defined by
 \begin{align}\label{Tg_u1g}
        {\cal T}_g(\bs A_\perp, \bs B_\perp)=\frac{(\bs A_\perp \cdot \bs B_\perp)^2+ \frac{\epsilon_f^2}{2}(\bs A_\perp^2+\bs B_\perp^2+\epsilon_f^2)
}{
(\epsilon_f^2+\bs A^2_\perp)(\epsilon_f^2+\bs B^2_\perp)}~.
       \end{align}
Summing all the contributions, we obtain the unpolarized fracture function for producing a real gluon with momentum fraction $\xi$ and transverse momentum $\bs p_{g\perp}$:    \begin{align}\label{eq:ug_final}
     xu_{1}^g(x,\xi,\bs p_{g\perp}^2)
&=\frac{4(N_c^2-1)}{\pi}
\int d^2\bs k_{g\perp} \left. \bigg[  {\cal T}_g(\bs k_{g\perp}-\bs p_{g\perp}, \bs k_{g\perp}-\bs p_{g\perp})+{\cal T}_g(-\bs p_{g\perp}, -\bs p_{g\perp}) \right.
 \\
&- \left. 2{\cal T}_g(\bs k_{g\perp}-\bs p_{g\perp}, -\bs p_{g\perp})
-\frac{1  }{\bs p_{g\perp}^2}\frac{\bs k_{g\perp }\cdot(\bs k_{g\perp}-\bs p_{g})_\perp(\bs p_{g\perp}^2-\bs k_{g\perp}^2)-\bs k_{g\perp}^2 \epsilon_f^2}{\epsilon_f^2+(\bs k_{g\perp}-\bs p_{g\perp})^2} \bigg] \right. {\cal F}_{x_g}(\bs k_{g\perp})
~.
\notag
\end{align}
The corresponding fracture function for producing a hadron $h$ can then be obtained by convoluting with the fragmentation function:
\begin{align}
    xu_{1}^g(x, \xi_h, \bs P_{h\perp}^2) \Big \vert_{h}= \int\frac{d z}{z^2} \,  d_{h/g}(z) \, xu_{1}^g\left(x,\frac{\xi_h}{z},\frac{\bs P_{h\perp}^2}{z^2}\right)\bigg\vert_{g} ,
\end{align}
where $d_{h / g}(z)$ is the gluonic fragmentation function into hadron $h$.

Using the energy sum rule in Eq.~\eqref{eq:sumrule_f1} together with the momentum sum rule for fragmentation functions (FF),
\begin{align}
    \sum_h \int_0^1 d z \,  z d_{h / a}(z)=1~,
\end{align}
we find that the unpolarized gluon NEC can be expressed in terms of the adjoint dipole gluon distributions as
\begin{align}
   &xf_{1}^g(x,\theta)=\frac{2(N_c^2-1)}{\theta^2(2\pi)^4}
\int_0^{1-x}d \xi
\int d^2\bs k_{g\perp} \, \bs p_\perp^2 \, \bigg[ {\cal T}_g(\bs k_{g\perp}-\bs p_\perp, \bs k_{g\perp}-\bs p_\perp)+{\cal T}_g(-\bs p_\perp, -\bs p_\perp)
\notag \\
&\;\;\;-2{\cal T}_g(\bs k_{g\perp}-\bs p_\perp, -\bs p_\perp)
-\frac{1}{\bs p_\perp^2}\frac{\bs k_{g\perp }\cdot(\bs k_{g\perp}-\bs p)_\perp(\bs p_\perp^2-\bs k_{g\perp}^2)-\bs k_{g\perp}^2 \epsilon_f^2}{\epsilon_f^2+(\bs k_{g\perp}-\bs p_\perp)^2}\bigg ]\, {\cal F}_{x_g}(\bs k_{g\perp}^2)\,\theta(x_g -\xi)~.
\label{eq:num_gNEC1}
\end{align}
where $ \bs p_\perp\equiv \frac{\theta\xi P^+}{\sqrt{2}} \bs n_t $ is the transverse momentum of final-state gluon that initiates the measured energy flux within the solid angle $(\theta,\phi)$. Following Ref.~\cite{Liu:2023aqb}, here we have imposed a kinematic constraint, $\xi<x_g$,  in order to ensure that the outgoing gluon momentum, $\xi P^+$, does not exceed the parent gluon momentum, $x_g P^+$.

\subsubsection{Linearly polarized gluon fracture function and NEC}

 Similar to the unpolarized gluon case, we deduce from decomposition~\eqref{eq:gluonFrF} the following projection onto the linearly polarized fracture function,
        \begin{align}
     \frac{t_{1}^{h,g}}{2M^2} =&\frac{2}{ (\bs p_{g\perp}^2)^2}\Big( p_{g\perp}^\alpha   p_{g\perp}^\beta + \frac{1}{2}g_\perp^{\alpha \beta} \bs p_{g\perp}^2\Big){\cal M}_{\text{FrF},\alpha \beta}~. \label{eq:pro2}
       \end{align}
Unlike the unpolarized case, only the interference diagram in Fig.~\ref{fig:gluonNEEC}, corresponding to Eq.~\eqref{eq:diagramC}, produces linearly polarized gluons. This yields
           \begin{align}
    \frac{xt_{1}^{h,g}(x,\xi,\bs p_{g\perp}^2)}{2M^2}
  = & - \frac{4(N_c^2-1)}{\pi} \frac{2}{(\bs p_{g\perp}^2)^3} \int d^2\bs k_{g\perp}\frac{  {\cal F}_{x_g}(\bs k_{g\perp})}{\epsilon_f^2+(\bs k_{g\perp}- \bs p_{g})_\perp^2} \bigg\{-\Big[2(\bs k_{g\perp}\cdot \bs p_{g\perp})^2-\bs k_{g\perp}^2 \bs p_{g\perp}^2\Big]  \frac{\epsilon_f^2}{2}
  \notag \\
  &+\Big[2(\bs k_{g\perp}\cdot \bs p_{g\perp})^2-\bs k_{g\perp}^2 \bs p_{g\perp}^2-\bs p_{g\perp}^2(\bs p_{g\perp} \cdot \bs k_g)\Big]\frac{1}{2}\Big[
\epsilon_f^2+(\bs k_{1g}- \bs p_{g\perp})^2+\bs p_{g\perp}^2-\bs k_{g\perp}^2\Big]\bigg\}
~.
\label{eq:tg_final}
 \end{align}
 Then, the corresponding fracture function for producing a hadron $h$ can be obtained by
\begin{align}
 xt_{1}^{h,g} (x, \xi_h, \bs P_{h\perp}^2)\Big \vert_{h}= \int\frac{d z}{z^2}   d_{h/g}(z)  xt_{1}^{h,g} \big(x,\frac{\xi_h}{z},\frac{\bs P_{h\perp}^2}{z^2}\big)\Big \vert_{g}
\end{align}
where $d_{h / g}(z)$ represents the gluonic FF into hadron $h$.

With the energy sum rule in Eq.~\eqref{eq:sumrule_h1} and the momentum sum rule of FFs, the linearly polarized gluon NEC becomes  \begin{align}
   xh_{1}^{t,g}(x,\theta)&=-\frac{2(N_c^2-1)}{\theta^2(2\pi)^4}
\int_0^{1-x}d \xi\int d^2\bs k_{g\perp} \frac{{\cal F}_{x_g}(\bs k_{g\perp}^2)}{\epsilon_f^2+(\bs k_{g\perp}- \bs p)_\perp^2} \bigg\{-\Big[2\frac{(\bs k_{g\perp}\cdot \bs p_\perp)^2}{\bs p_\perp^2}  -\bs k_{g\perp}^2 \Big] \epsilon_f^2
  \notag \\
  & +\Big[2\frac{(\bs k_{g\perp}\cdot \bs p_\perp)^2}{\bs p_\perp^2}-\bs k_{g\perp}^2-\bs p_\perp \cdot \bs k_{g\perp})]\Big]\Big[
\epsilon_f^2+(\bs k_{g}- \bs p)_\perp^2+\bs p_{\perp}^2-\bs k_{g\perp}^2\Big]
 \bigg\}\theta(x_g -\xi)~,
  \label{eq:num_gNEC2}
\end{align}
 where $ \bs p_\perp\equiv \frac{\theta\xi P^+}{\sqrt{2}} \bs n_t $. Here, the kinematic constraint $\xi<x_g$ is imposed similarly to Eq.~\eqref{eq:num_gNEC1}

  \subsubsection{Discussion: the dilute limit and collinear matching }
  \label{sec:dilute_limit}

  It is instructive to study the above CGC results in the dilute limit and verify their matching onto the collinear formalism. This limit corresponds to the large $\bs p_{g\perp}$ region of the fracture functions, which is equivalent to the large $\theta$ region of the NECs.

   Let us begin with the fracture functions. In the large-$\bs p_{g\perp}$ region, where $\bs k_{g\perp} \ll\, \bs p_{g\perp}$, one can perform a Taylor expansion of the momentum-space impact factor:
\begin{align}
{\cal H}(\bs p_{g\perp},\bs k_{g\perp})&
={\cal H}\left(\bs p_{g\perp},\bs k_{g\perp}=0\right)+\frac{\partial {\cal H}\left(\bs p_{g\perp}, \bs k_{g\perp}\right)}{\partial \bs k_{g\perp}^\alpha}\Bigg\vert_{\bs k_{g\perp}=0} \bs k_{g\perp}^\alpha+\frac{1}{2} \frac{\partial^2 {\cal H}\left(\bs p_{g\perp}, \bs k_{g\perp}\right)}{\partial \bs k_{g\perp}^\alpha \partial \bs k_{g\perp}^\beta}\Bigg\vert_{\bs k_{g\perp}=0} \bs k_{g\perp}^\alpha \bs k_{g\perp}^\beta
+\cdots.
\end{align}
For $u_{1}^{g}$ and $t_{1}^{h,g}$ given in Eqs.~\eqref{eq:ug_final} and \eqref{eq:tg_final}, the first two terms appear to vanish as a consequence of the Ward identity. Therefore, the leading contribution arises from the third term quadratic in $\bs k_{g\perp}$. After carrying out the expansion, we obtain the following expressions in terms of the collinear gluon distribution $f_g(x)$:
\begin{align}
    & u_{1}^g(x,\xi,\bs p_{g\perp}^2)
=\frac{4g_s^2 N_c}{\bs p_{g\perp}^2} \frac{(\xi^2+x^2+\xi x)^2}{x(\xi+x)^4} \, x_g f(x_g)
~,
\notag \\
&\frac{t_{1}^{h,g}(x,\xi,\bs p_{g\perp}^2)}{2M^2}=\frac{4g_s^2 N_c}{(\bs p_{g\perp}^2)^2} \frac{\xi^2}{x(\xi+x)^2} \,  x_gf(x_g)~
~.
\label{eq:FrF_dilute}
\end{align}
Here, we have used the operator relation between the collinear gluon distribution $f_g(x)$ and the adjoint dipole gluon distribution ${\cal F}_{x_g}(\bs k_{g\perp})$:
\begin{align}
 x_g f_g(x_g)=\frac{N_c^2-1}{\pi g_s^2 N_c ~}\int d^2\bs k_{g\perp}  \bs k_{g\perp}^2 \,  {\cal F}_{x_g}(\bs k_{g\perp})~.
 \label{eq:connection_Gdipole}
\end{align}
If we further identify $x_g=\xi+x$ from momentum conservation, Eq.~\eqref{eq:FrF_dilute} reproduces the fracture functions computed in the collinear formalism in the small-$x$ limit; see our derivation in Appendix~\ref{appendix:collinear}. We find that the unpolarized gluon fracture function $u_{1}^g$ exhibits the same $1/\bs p_{g\perp}^2$ scaling behavior as the unpolarized quark fracture function derived in Refs.~\cite{Chen:2021vby,Chen:2024bpj,Caucal:2025qjg}, whereas the linearly polarized gluon fracture function $t_{1}^{h,g}$ is additionally suppressed by a factor of $M^2/\bs p_{g\perp}^2$. This extra suppression is purely kinematical and originates from the projection or normalization, as seen from Eq.~\eqref{eq:pro2}.

With the dilute limit available for the fracture functions, we next derive the corresponding large-$\theta$ limit of the gluon NECs $f_1^{g}$ and $h_1^{t,g}$ by expanding in the region $ \bs k_{g\perp}\ll \bs p_{g\perp}= \frac{\theta\xi P^+}{\sqrt{2}} \bs n_t $ in Eqs.~\eqref{eq:num_gNEC1} and \eqref{eq:num_gNEC2}. Introducing the variable $z$ through $\xi=\frac{1-z}{z}x$, we obtain the following large-$\theta$ behavior of the gluon NECs in the dilute limit:
\begin{align}
f_{1}^g(x,  \theta)=&\frac{\alpha_s}{4 \pi^2} \frac{1}{\theta^2} \int_x^1 \frac{d z}{z} \, P_{g g}(z)\left(1-z\right) x_gf_g\left(x_g\right)~,
\notag \\
  h_{1}^{t,g}(x,\theta)
  =&\frac{\alpha_s}{4\pi^2}\frac{1}{\theta^2}\int^1_x \frac{dz}{z} \,
\tilde P_{g g}(z)\left(1-z\right)x_gf_g(x_g)~,
\label{eq:NEC_dilute}
\end{align}
with $ P_{g g}(z)$ and $\tilde P_{g g}(z)$ the collinear splitting kernels from unpolarized and linearly polarized gluons, respectively~\cite{Nadolsky:2007ba,Catani:2010pd}
:
\begin{align}
  P_{g g}(z)=2 C_A \frac{(1-z+z^2)^2}{z(1-z)}~,~\tilde P_{g g}(z) =\frac{2C_A(1-z)}{z}~.
\end{align}
Beyond the small-$x$ region, one can apply the momentum conservation and identify $x_g=\xi+x=x/z$. The common factor $(1-z)$ in Eq.~\eqref{eq:NEC_dilute} then represents the energy weight of the emitted parton entering the detector in the  TFR. The above results reproduce the gluon NECs generated by collinear gluon splittings, as given in Ref.~\cite{Guo:2024jch}. This shows that both $f_{1}^g$ and $h_{1}^{t,g}$ obey the same $1/\theta^2$ scaling as the unpolarized quark NEC~\cite{Liu:2022wop,Chen:2024bpj}.

\subsubsection{Discussion: Fracture function for jet production}
In Ref.~\cite{Caucal:2025qjg}, the unpolarized gluon \emph{jet fracture function} was derived from single-inclusive jet production in DIS by taking the TFR limit.  We notice that this quantity can be identified with the phase-space integral over the momentum fraction $\xi$ of the \emph{unpolarized gluon fracture function} $u_1^g(x,\xi,\bs p_{\perp}^2)$, defined in Eq.~\eqref{eq:pro1} and given explicitly in Eqs.~\eqref{Tg_u1g} and \eqref{eq:ug_final}:
\begin{align}
  u_1^g(x,\bs p_{\perp})\bigg\vert_{\text{Jet}}=  \int\frac{d\xi}{2\xi(2\pi)^3} \, u_1^g(x,\xi,\bs p_{g\perp}) = -g_{\perp}^{\alpha\beta}\int \frac{d\xi}{2\xi(2\pi)^3} \, {\cal M}_{\text{FrF},\alpha \beta}(x,\xi,\bs p_{g\perp})~.
  \label{eq:jetFrF}
\end{align}A direct comparison with Ref.~\cite{Caucal:2025qjg}, however, is not straightforward because specific kinematic cutoffs are imposed for  different diagrams at intermediate stages of that calculation. Performing the indefinite integral over $\xi$ in Eq.~\eqref{eq:jetFrF}, we obtain
\begin{align}
    &\int\frac{d\xi}{2\xi} \, xu_1^g(x,\xi,\bs p_{g\perp})     \\
    &=  \frac{4(N_c^2-1)}{2\pi} \int d^2\bs k_{g\perp} \, {\cal F}_{x_g}(\bs k_{g\perp})\left\{  \frac{ \bs k_{g\perp}^2}{\bs p_{g\perp}^2} \ln|\xi| + \frac{x}{2(\xi+x)} + \frac{x\bs p_{g\perp}^2}{2\left[(\bs p_{g\perp} - \bs k_{g\perp})^2\xi + x\bs p_{g\perp}^2\right]}     \right. \notag \\
    &\;\;\;\;+ \frac{\bs k_{g\perp}^2(\bs p_{g\perp} - \bs k_{g\perp})^2 - \bs p_{g\perp}^2(\bs p_{g\perp} - \bs k_{g\perp})^2 + 3\bs k_{g\perp}^2\bs p_{g\perp}^2 - \bs p_{g\perp}^4 - \bs k_{g\perp}^4 }{2\bs p_{g\perp}^2\left[\bs p_{g\perp}^2 - (\bs p_{g\perp} - \bs k_{g\perp})^2\right]}  \ln\left|2(\bs p_{g\perp} - \bs k_{g\perp})^2\xi + 2x\bs p_{g\perp}^2 \right|     \notag \\
    &\;\;\;\;+ \left. \frac{ 4\left[\bs p_{g\perp}\cdot(\bs p_{g\perp} - \bs k_{g\perp})\right]^2 - 3\bs p_{g\perp}^2 (\bs p_{g\perp} - \bs k_{g\perp})^2 + \bs p_{g\perp}^4}{2\bs p_{g\perp}^2\left[\bs p_{g\perp}^2 - (\bs p_{g\perp} - \bs k_{g\perp})^2\right]}   \ln\left|2(\xi+x)\bs p_{g\perp}^2\right| \right\} + C \, ,      \notag
\end{align}
where we noticed that the terms purely linear in $\bs k_{g\perp}$ vanish upon integration over $\bs k_{g\perp}$. Here, $C$ is the constant term resulting from an indefinite integral. To proceed, we impose the integration limit, $x \bs p_{g\perp}^2/Q^2\leq\xi\leq 1$, employed in~\cite{Caucal:2025qjg}\footnote{This corresponds to $z_0\leq z\leq 1$ in the notation of~\cite{Caucal:2025qjg}.}. In addition, a kinematic constraint of $\xi\geq x$, corresponding to $z\leq \bs P_{\perp}^2/Q^2$ in the notation of~\cite{Caucal:2025qjg}, is imposed on diagram B of Fig. \ref{fig:gluonNEEC}, in which the gluons are emitted after the shockwave in both the amplitude and the complex conjugate\footnote{In~\cite{Caucal:2025qjg}, this additional constraint results from the approximation used to simplify the expressions for the NLO-0 and NLO-3 terms.}. As a result, we obtain the unpolarized gluon jet fracture function of the form
\begin{align}\label{FaridInt_FrF_full}
    &\int\limits_{x\bs p_{g\perp}^2/Q^2}^1 \frac{d\xi}{2\xi} \, xu_1^g(x,\xi,\bs p_{g\perp})\bigg|_{\text{(A)+(B)+(C)}}~-\int\limits_{x\bs p_{g\perp}^2/Q^2}^x \frac{d\xi}{2\xi} \, xu_1^g(x,\xi,\bs p_{g\perp}) \bigg|_{\text{(B)}}
    \\
    &= \frac{4(N_c^2-1)}{2\pi} \int d^2\bs k_{g\perp} \, {\cal F}_{x_g}(\bs k_{g\perp})\left\{  \frac{ \bs k_{g\perp}^2}{\bs p_{g\perp}^2} \ln\frac{1}{x} + \frac{x}{2(1+x)} - \frac{Q^2}{2\left[\bs p_{g\perp}^2+Q^2\right]} + \frac{x\bs p_{g\perp}^2}{2\left[(\bs p_{g\perp} - \bs k_{g\perp})^2 + x\bs p_{g\perp}^2\right]}     \right. \notag \\
    &\;\;\;\;\;\;\;\; - \frac{Q^2}{2\left[(\bs p_{g\perp} - \bs k_{g\perp})^2 + Q^2\right]}   + \frac{ 4\left[\bs p_{g\perp}\cdot(\bs p_{g\perp} - \bs k_{g\perp})\right]^2 - 3\bs p_{g\perp}^2 (\bs p_{g\perp} - \bs k_{g\perp})^2 + \bs p_{g\perp}^4}{2\bs p_{g\perp}^2\left[\bs p_{g\perp}^2 - (\bs p_{g\perp} - \bs k_{g\perp})^2\right]}   \ln\left[\frac{1+x}{\left(1+\frac{\bs p_{g\perp}^2}{Q^2}\right)x}\right]  \notag \\
    &\;\;\;\;\;\;\;\;+ \left. \frac{\bs k_{g\perp}^2(\bs p_{g\perp} - \bs k_{g\perp})^2 - \bs p_{g\perp}^2(\bs p_{g\perp} - \bs k_{g\perp})^2 + 3\bs k_{g\perp}^2\bs p_{g\perp}^2 - \bs p_{g\perp}^4 - \bs k_{g\perp}^4 }{2\bs p_{g\perp}^2\left[\bs p_{g\perp}^2 - (\bs p_{g\perp} - \bs k_{g\perp})^2\right]}  \ln\left[\frac{(\bs p_{g\perp} - \bs k_{g\perp})^2 + x\bs p_{g\perp}^2}{\left((\bs p_{g\perp} - \bs k_{g\perp})^2\frac{\bs p_{g\perp}^2}{Q^2} + \bs p_{g\perp}^2\right)x} \right] \right\}  .      \notag
\end{align}
If we take the limit of~\eqref{FaridInt_FrF_full} where $x\ll 1$ and $\bs p_{g\perp}^2 \ll \bs k_{g\perp}^2 \ll Q^2$, then we arrive at the expression,
\begin{align}\label{FaridInt_FrF_approx}
    &\frac{4(N_c^2-1)}{2\pi} \int d^2\bs k_{g\perp} \, {\cal F}_{x_g}(\bs k_{g\perp})\left\{  \frac{ \bs k_{g\perp}^2}{\bs p_{g\perp}^2} \ln\frac{1}{x} - 1  +  \frac{4(\bs p_{g\perp}\cdot\bs k_{g\perp})^2  - 3\bs p_{g\perp}^2\bs k_{g\perp}^2}{2\bs p_{g\perp}^2\bs k_{g\perp}^2}   \ln\left[\frac{(\bs p_{g\perp} - \bs k_{g\perp})^2}{\bs p_{g\perp}^2} \right] \right\}  .
\end{align}
The finite term and the term with rapidity logarithm match their counterparts in Eq.~(79) of~\cite{Caucal:2025qjg}. On the other hand, the term with transverse logarithm in Eq.~(79) of~\cite{Caucal:2025qjg} simplifies under the $\bs p_{g\perp}^2 \ll \bs k_{g\perp}^2 \ll Q^2$ limit to
\begin{align}
    &\left[\left(1 - \frac{2\left[\bs p_{g\perp}\cdot(\bs p_{g\perp} - \bs k_{g\perp})\right]^2}{\bs p_{g\perp}^2(\bs p_{g\perp} - \bs k_{g\perp})^2}\right)\frac{(\bs p_{g\perp} - \bs k_{g\perp})^2}{\bs p_{g\perp}^2 - (\bs p_{g\perp} - \bs k_{g\perp})^2} - \frac{(\bs p_{g\perp} - \bs k_{g\perp})^2 - \bs k_{g\perp}^2}{2\bs p_{g\perp}^2}\right]\ln\left[\frac{(\bs p_{g\perp} - \bs k_{g\perp})^2}{\bs p_{g\perp}^2} \right] \notag \\
    &\simeq \frac{4(\bs p_{g\perp}\cdot\bs k_{g\perp})^2  - 3\bs p_{g\perp}^2\bs k_{g\perp}^2}{2\bs p_{g\perp}^2\bs k_{g\perp}^2} \ln\left[\frac{(\bs p_{g\perp} - \bs k_{g\perp})^2}{\bs p_{g\perp}^2} \right] ,
\end{align}
which also agrees with the corresponding term in Eq.~\eqref{FaridInt_FrF_approx}. Hence, the calculation performed in this work achieves complete agreement with~\cite{Caucal:2025qjg} for the unpolarized gluon jet fracture function in the overlapping $x\ll 1$ and $\bs p_{g\perp}^2 \ll \bs k_{g\perp}^2 \ll Q^2$ limits.

\section{Linearly polarized gluonic NEC as a probe for saturation effects}\label{sec:numerics}

In this section, we show the linearly polarized gluon NEC as a novel and sensitive probe of saturation effects in DIS at small-$x_B$. The key predicted signature is a nuclear suppression of a $\cos 2\phi$ azimuthal asymmetry in the energy pattern measured in the TFR of DIS. We begin by introducing the corresponding energy-pattern observable and its factorization in the TFR in the Bjorken limit. We then perform a numerical study of this observable for the EIC kinematics.
\subsection{Observable: the $\cos2\phi$ azimuthal asymmetry of the DIS energy pattern}
\begin{figure}[t]
  \centering
\includegraphics[scale=0.2]{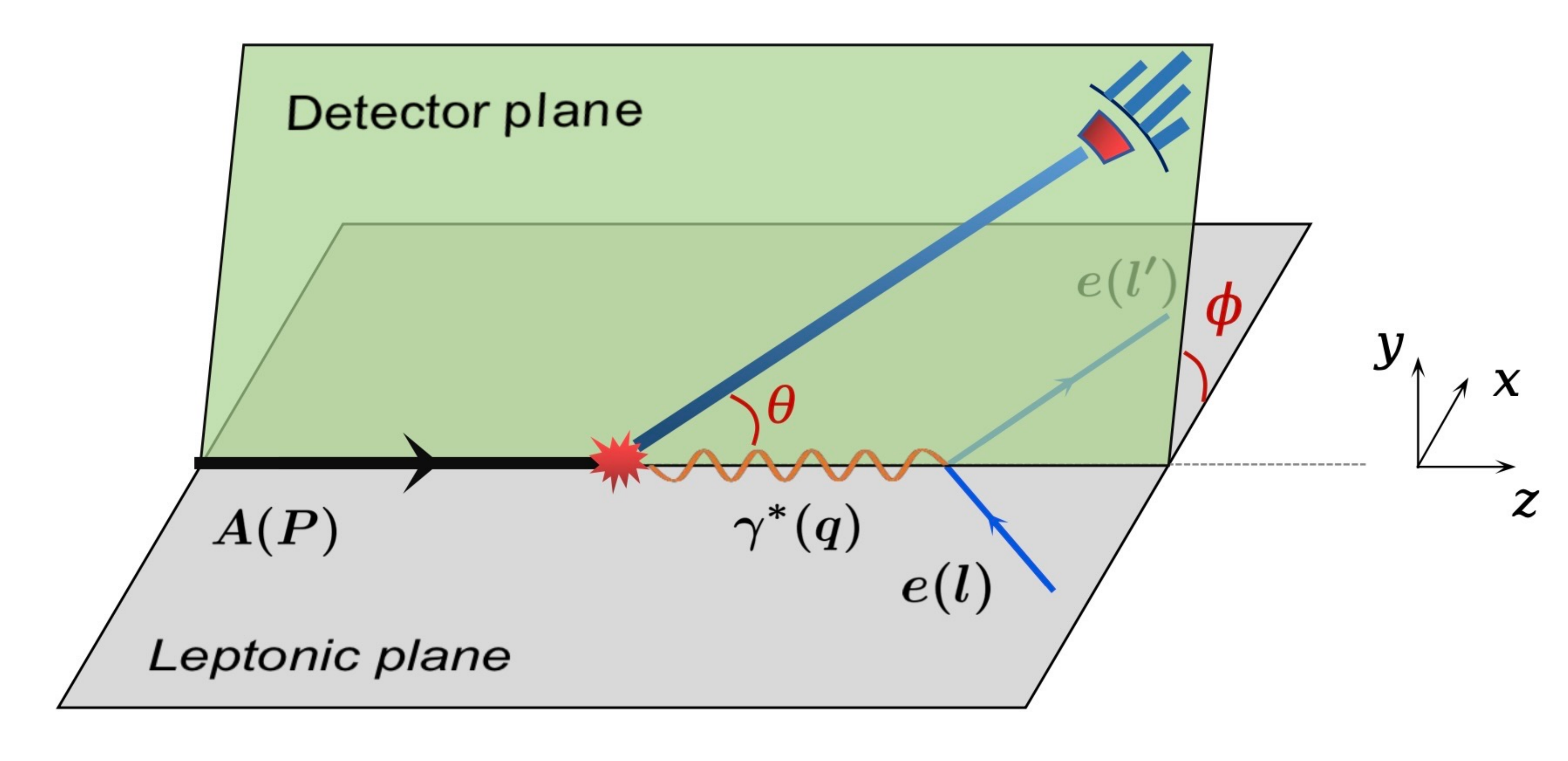}

\caption{Illustration of the DIS energy pattern in the Breit frame.}
\label{fig:frame}
\end{figure}

We consider measuring the angular distribution of an  energy flux in the target fragmentation region in DIS. The associated energy pattern cross section is defined as~\cite{Liu:2022wop,Chen:2024bpj}
\begin{align}
\Sigma(\theta, \phi)=\sum_{h} \int \mathrm{~d} \sigma^{l+p \rightarrow l^{\prime}+h+X} \frac{E_h}{E_N} \delta\left(\theta^2-\theta_h^2\right) \delta\left(\phi-\phi_h\right)
\end{align}
where $\mathrm{~d} \sigma^{l+p \rightarrow l^{\prime}+h+X}$ represents the differential cross section of single-hadron inclusive DIS, and $\sum_h$ denotes the inclusive sum over the hadron species.
As shown in Fig.~\ref{fig:frame}, we work in the Breit frame, where the incoming proton is along the $z$ direction, and the photon momentum is all in the $-z$ direction:
\begin{align}
P^\mu= &\Big(\frac{Q}{x_B\sqrt{2}},0,\bs 0_\perp\Big)~,~~q^\mu=\Big(-\frac{Q}{\sqrt{2}},\frac{Q}{\sqrt{2}},\bs 0_\perp\Big)~.
\end{align}
The polar angle $\theta$ is measured with respect to the nucleon beam, and the azimuthal angle $\phi$ is defined relative to the lepton scattering plane. The $x$- and $y$-axes are chosen such that the incoming lepton has the transverse momentum $\bs l^\mu_\perp=( Q\sqrt{1-y}/{y},0)
$. The standard DIS kinematic variables are defined by
\begin{align}
Q^2=-q^2~,\quad x_B=\frac{Q^2}{2P\cdot q}~,\quad y=\frac{Q^2}{s x_{\text{B}}} ~
\end{align}
with $\sqrt{s}$ as the center-of-mass energy of the $ep$ collision.

To extract the effects of linearly polarized gluons, we consider the $\cos 2\phi$ asymmetry defined as \begin{align}
  \langle \cos 2\phi \rangle=  2 \frac{\int_0^{2\pi} d\phi \cos2\phi\Sigma(\theta, \phi)}{\int_0^{2\pi} d\phi \Sigma(\theta, \phi)}~.  \label{eq:asym}
\end{align}
This asymmetry characterizes an elliptic anisotropy of the energy pattern with respect to the lepton scattering plane specified in Fig.~\ref{fig:frame}, which is namely a difference between energy deposition along and perpendicular to that plane. In the one-photon-exchange approximation, the energy pattern relevant to this asymmetry is parameterized in terms of the following structure functions~\cite{Chen:2024bpj}:
\begin{align}
&\frac{d \Sigma(\theta, \phi)}{d x_B d Q^2 }
=\frac{2\alpha_\mathrm{em}^2}{x_B Q^4} \bigg[A(y) \Sigma_{U U,T}+B(y) \Sigma_{U U,L}+B(y)\Sigma_{UU}^{\cos2\phi} \cos2\phi
\bigg]~,
\label{eq:SF}
\end{align}
where $A(y)=1- y+y^2/2$ and $B(y)=1-y$. The structure functions are scalar functions of $x_B,~Q^2$ and the polar angle $\theta$. Here,
  $\Sigma_{U U,T}$ and $ \Sigma_{U U,L}$ denote the azimuthally independent structure functions associated with transverse and longitudinal photons, respectively, while $\Sigma_{UU}^{\cos2\phi}$ is the structure function associated with the $\cos 2\phi$ modulation of primary interest here. In terms of these structure functions, the asymmetry at fixed $(x_B,Q^2)$ is given by
\begin{align}
 \langle \cos 2\phi \rangle=\frac{B(y) \Sigma_{U U}^{\cos 2\phi}}{A(y) \Sigma_{U U,T} }~,
 \label{eq:asym2}
\end{align}
where we retain only the transverse-photon contribution in the normalization, since the $\cos 2\phi$ azimuthal asymmetry can be generated only by a transverse virtual photon~\cite{Chen:2024brp}. Furthermore, as we show below, $\Sigma_{UU,L}$ vanishes relative to $\Sigma_{UU,T}$ when one restricts to the leading-order contribution in $\alpha_s$.

\subsection{Factorization with NECs in the TFR }
To make explicit the connection between the energy-pattern asymmetry $\langle\cos 2 \phi\rangle$ and the linearly polarized gluon NECs, we consider the DIS energy pattern in the target fragmentation region, $\theta \ll \pi$, in the Bjorken limit $Q \gg \Lambda_{\mathrm{QCD}}$.
In this kinematic regime, the observed energy flux mainly arises from the fragmentation of the target remnant after the virtual photon removes a parton from the nucleon. Because the photon virtuality $Q$ is much larger than the nonperturbative scale, the energy flux and the associated long-distance target–fragmentation dynamics are well separated from the short-distance partonic scattering, while still retaining correlations with the momentum and polarization of the struck parton. These correlations are nonperturbative in nature and are encoded in NECs.

Therefore, the DIS energy-pattern cross section in the TFR can be factorized into quark and gluon NECs convoluted with perturbatively calculable hard kernels. At leading power of $1/Q$, it takes the form~\cite{Liu:2022wop}:
\begin{align}
	 \Sigma(x_B,\theta)=\sum_{a={q,g}}\int^1_{x_B} \frac{d z}{z}  H_a\Big(\frac{x_B}{z},\frac{Q}{\mu}\Big){\cal M}_{\text{NEC}}^{a}(z,\theta,\mu)
\end{align}
where $H_{a}$ represent the hard scattering part for parton $a$, and the NECs  ${\cal M}_{\text{NEC}}^{a}$ obey the DGLAP evolution equation~\cite{Cao:2023oef,Chen:2024bpj,Gao:2025cwy}, with the factorization scale denoted as $\mu$.
As shown in Ref.~\cite{Chen:2024bpj}, the TFR factorization formula for the DIS energy pattern can be rigorously derived from the TFR factorization theorem for single-hadron inclusive DIS, originally established by Collins in the seminal work~\cite{Collins:1997sr}. The complete contributions to the structure functions have been derived out to subleading power in $1/Q$, in Ref.~\cite{Chen:2024bpj} by exploiting the connection to SIDIS and fracture functions.

An important observation is that the hard partonic subprocesses relevant to the DIS energy pattern in the target fragmentation region are described by the same perturbative diagrams as those in ordinary inclusive DIS. As a result, the hard kernels associated with the structure functions $\Sigma_{U U, T}$ and $\Sigma_{U U, L}$ can be obtained directly from the corresponding kernels for the inclusive DIS structure functions $F_T$ and $F_L$. In particular, for a transversely polarized virtual photon, the leading contribution in $\alpha_s$ is generated by the unpolarized quark NEC,
\begin{align}
	\Sigma_{U U, T}=\sum_{q, \bar{q}} e_q^2 x_B f_1^q\left(x_B, \theta\right)+\mathcal{O}\left(\alpha_s\right) .
	  \label{eq:UUT}
\end{align}
For a longitudinally polarized virtual photon, the first nonvanishing contribution arises at $\mathcal{O}\left(\alpha_s\right)$~\cite{Chen:2024bpj}:
\begin{align}
   \Sigma_{UU,L} = \frac{\alpha_s}{2\pi}  \sum_{q,\bar q} e_q^2 \int^1_{x_B} \frac{d z}{z} \bigg[
 4T_F \Big(\frac{x_B}{z}\Big)^2\Big(1-\frac{x_B}{z}\Big) zf_{1}^g(z,\theta) +2 C_F \Big(\frac{x_B}{z}\Big)^2 zf_{1}^q(z,\theta) \bigg]+{\cal O}(\alpha_s^2)~.
  \label{eq:UUL}
\end{align}
 where both the unpolarized quark and gluon NECs contribute. Similar to the case of $F_L$, at small $x_B$, $\Sigma_{UU,L}$ can be approximated as~\begin{align}
	  \Sigma_{UU,L}
 =&\frac{\alpha_s}{2\pi}
 T_F\frac{4}{6} \sum_{q,\bar q} e_q^2 x_Bf_{1}^g(x_B,\theta) +\frac{\alpha_s}{2\pi}  C_F  \sum_{q,\bar q} e_q^2 x_Bf_{1}^q(x_B,\theta) ~,
\end{align}
which is also recently confirmed by an explicit CGC calculation in Ref.~\cite{Caucal:2025qjg}.

However, unlike ordinary collinear PDFs involved in inclusive DIS, NECs provide access to linearly polarized gluons even in an unpolarized nucleon~\cite{Li:2023gkh}. This is possible because of the azimuthal correlation between gluon linear polarization and the energy flux measured in the target fragmentation region. More explicitly, the relevant component in the gluonic NEC matrix can be written as\begin{align}
	{\cal M}_{\text{NEC}}^{\alpha \beta}\propto &-
\Big( n_{t}^\alpha  n_{t}^\beta +\frac{1}{2}g_\perp^{\alpha \beta} \Big) h_{1}^{t,g}=\frac{1}{2}\left(\begin{array}{cc}
\cos \left(2 \phi_t\right) & \sin \left(2 \phi_t\right) \\
\sin \left(2 \phi_t\right) & -\cos \left(2 \phi_t\right)
\end{array}\right)h_{1}^{t,g}~,
\end{align}
where $\phi_t$ is the azimuthal angle of the measured energy flux in the transverse plane.

When these linearly polarized gluons enter the hard scattering mediated by a transverse photon, they manifest as a $\cos 2 \phi$ asymmetry of the TFR energy flux relative to the lepton scattering plane. At leading order, the relevant hard scattering arises from the interaction of the gluon with the quark-antiquark pair produced by the virtual photon. The corresponding structure function factorizes in terms of the linearly polarized gluon NEC as~\cite{Chen:2024bpj,Chen:2024brp}
\begin{align}
& \Sigma_{UU}^{\cos 2\phi} =  - \frac{\alpha_s }{2\pi}  \sum_{q,\bar q} e_q^2
\int^1_{x_B} \frac{d z}{z} T_F \Big(\frac{x_B}{z}\Big)^3 z h^{t,g}_{1}(z,\theta)+{\cal O}(\alpha_s^2)~,
\label{eq:cos2phi}
\end{align}
where $T_F=1/2$.

This provides a novel and sensitive probe of linearly polarized gluons through NECs~(see also~\cite{Cao:2023oef,Collins:1997sr}). In comparison with the conventional TMD approach using two-particle correlations~(e.g., dijet or di-hadron~\cite{Boer:2010zf, Metz:2011wb,Dominguez:2011br, Pisano:2013cya, Dumitru:2015gaa,Boer:2016fqd,Hatta:2020bgy,Hatta:2021jcd}), the NEC framework offers several advantages~\cite{Li:2023gkh,Guo:2024jch,Guo:2024vpe}. First, as shown in Eq.~\eqref{eq:cos2phi}, the $\cos 2 \phi$ asymmetry in the NEC approach is formulated in collinear factorization, where soft-gluon effects cancel after the inclusive sum over unobserved hadrons. Thus, this observable is free from complications caused by Sudakov logarithms, which require resummation in the TMD approach~\cite{Sun:2011iw,Gutierrez-Reyes:2019rug} and lead to suppression in the nonperturbative region (see e.g.,~\cite{Boer:2017xpy,Marquet:2025jdr}).  Moreover, in the TMD framework, the corresponding $\cos 2\phi$ asymmetry can receive sizable contributions from soft-gluon radiation unrelated to the gluon polarization of the target, thereby contaminating the signal~\cite{Hatta:2020bgy,Hatta:2021jcd,Marquet:2025jdr,Gao:2026azd}. In addition, the TMD approach typically relies on dijet or dihadron measurements, which require either jet reconstruction algorithms or tracking detectors. By contrast, the NEC approach is based on the inclusive measurement of a single energy flux and therefore only requires recording the deposited energy in the calorimeter.

Given these advantages, we now investigate the $\cos 2\phi$ asymmetry in the small-$x_B$ region and explain why it can serve as a promising probe of saturation effects. First, as in the case of gluon TMDs, the linearly polarized gluon NEC is expected to be enhanced at small $x$. While this enhancement is already suggested by the splitting kernel at large $\theta$ in Eq.~\eqref{eq:NEC_dilute}, it is natural to expect that it persists into the small-$\theta$ region with $\theta Q \lesssim Q_s$, where saturation dynamics becomes important. In the small-$x_B$ limit, we therefore approximate the convolution integral in Eq.~\eqref{eq:cos2phi} by replacing $  zh_1^{t,g}(z,\theta) \to x_Bh_1^{t,g}(x_B,\theta)$.
With this approximation, the corresponding structure function becomes
\begin{align}
\Sigma_{UU}^{\cos 2\phi}
=
-\frac{\alpha_s}{2\pi}\frac{T_F}{3}\sum_{q,\bar q} e_q^2\, x_B\, h_1^{t,g}(x_B,\theta)\, .
\end{align}
Thus, at leading order in $\alpha_s$, the $\cos 2\phi$ asymmetry takes the form
\begin{align}
 \langle \cos 2\phi \rangle
 =
 -\frac{\alpha_s}{2\pi}\frac{T_F}{3}\frac{B(y)}{A(y)}
 \frac{\sum_{q,\bar q} e_q^2\, x_B\, h_1^{t,g}(x_B,\theta)}
 {\sum_{q,\bar q} e_q^2\, x_B\, f_1^q(x_B,\theta)}
 \, .
 \label{eq:cos2phi_num}
\end{align}
Remarkably, one finds that the asymmetry is determined by the ratio of the {\it linearly polarized gluon} NEC $h_1^{t,g}$ to the {\it unpolarized quark} NEC $f_1^q$. This is in contrast to the conventional $\cos 2\phi$ asymmetries in two-particle correlations, which are typically determined by the ratio of two {\it gluon} TMDs.

\begin{figure}[t]
  \centering

 \includegraphics[scale=0.46]{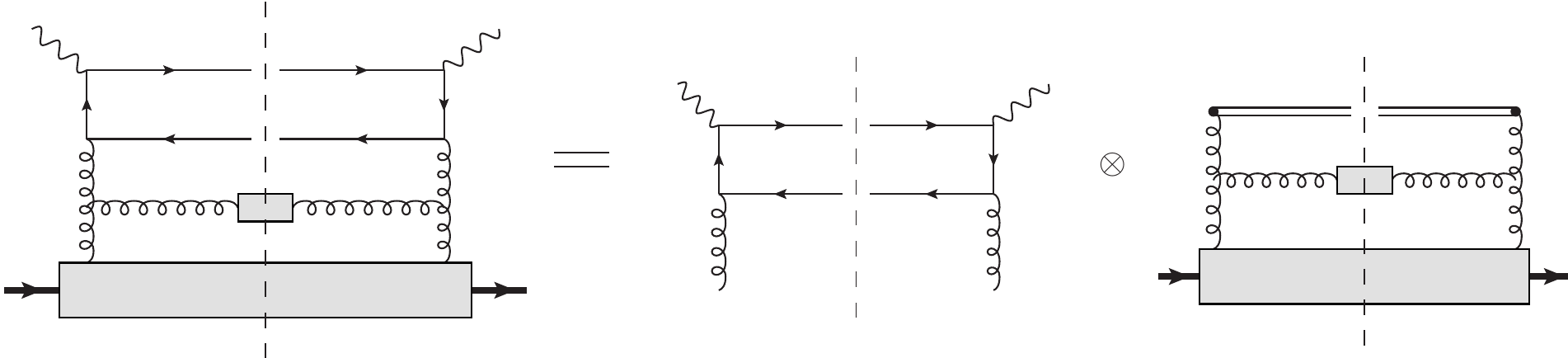}

\caption{Illustration of the TFR factorization of the $\cos 2\phi$ energy-pattern modulation starting from the full CGC description }
\label{fig:DIS_TFR}
\end{figure}

The form of the ratio in Eq.~\eqref{eq:cos2phi_num} makes the sensitivity of the $\cos 2\phi$ asymmetry to the saturation scale $Q_s$ at $\theta Q \lesssim Q_s$ manifest. This is because the small-$x$ gluonic NEC $h_1^{t,g}$ is determined by the gluon dipole $S$-matrix, as shown in Eq.~\eqref{eq:num_gNEC2}, whereas the small-$x$ unpolarized quark NEC $f_1^q$ is generated by the quark dipole $S$-matrix~\cite{Liu:2023aqb,Mantysaari:2025mht}. Since the gluon dipole can be approximated as the square of the quark dipole, see also Eq.~\eqref{eq:gluonquark}, the gluonic NEC $h_1^{t,g}$ is expected to be more sensitive to saturation effects than the quark NEC, especially in the region $\theta Q \lesssim Q_s$. With these results in hand, we turn in the next section to a numerical study of the sensitivity of the $\cos 2\phi$ asymmetry to saturation effects. In addition, an illustration of the TFR factorization of the $\cos 2\phi$ energy-pattern modulation from the CGC description is given in Fig.~\ref{fig:DIS_TFR}.

\subsection{Numerical prediction for the EIC}

We now present a numerical analysis of saturation effects for the gluon NECs $f_1^{g}$ and $h_1^{t,g}$ given in Eqs.~\eqref{eq:num_gNEC1} and \eqref{eq:num_gNEC2}, as well as for the $\cos 2\phi$ energy-pattern asymmetry in Eq.~\eqref{eq:cos2phi_num}. For predictions at the EIC, we consider a representative kinematic configuration:  $x_B=5\times 10^{-3}$, $Q^2=25~\mathrm{GeV}^2$, $\sqrt{s}=89~\mathrm{GeV}$.

The basic nonperturbative input for the gluonic NECs is the gluon dipole $S$-matrix defined in Eq.~\eqref{eq:gluondipole_coordinate}. For numerical estimates, we work in the large-$N_c$ limit, in which the gluon dipole $S$-matrix ${\cal S}_{x_g}$ can be expressed as the square of the quark dipole $S$-matrix ${S}_{x_g}$ defined in the fundamental representation. As a result, the adjoint-dipole gluon distribution in Eq.~\eqref{eq:gluondipole} can be written as
\begin{align}
 {\cal F}_{x_g}(\bs k_{g\perp})
=
\pi R^2
\int \frac{d^2\bs r_\perp}{(2\pi)^2}
\,e^{-i\bs r_\perp \cdot \bs k_{g\perp}}
\Big[ S_{x_g}(\bs r_\perp)\Big]^2~.
\label{eq:gluonquark}
\end{align}
Here we have assumed that the impact-parameter dependence of the target factorizes, with $\pi R^2$ representing the average transverse area of gluon distributions inside the target. The variable $x_g$ denotes the momentum fraction  that separates the fast-moving modes integrated out into the CGC effective theory from the active slow-moving partons.
In our calculations, we follow \cite{Liu:2023aqb} and choose $x_g=x_B$. To incorporate small-$x$ evolution effects, we obtain the quark dipole $S$-matrix ${ S}_{x_g}(\bs r_\perp)$ by solving the rcBK evolution equation~\cite{Albacete:2007yr, Albacete:2010sy, Balitsky:1995ub, Balitsky:2006wa, Balitsky:2007feb, Berger:2010sh, Kovchegov:2006wf, Kovchegov:1999yj, Kovchegov:2006vj, Gardi:2006rp, Golec-Biernat:2001dqn} using the modified McLerran--Venugopalan model~\cite{Albacete:2010sy} as the initial condition at $x_0=0.01$:
\begin{equation}
 S_{x_0=0.01}(\boldsymbol{r})
=
\exp\left[
-\frac{\left(Q_{s0}^2 \boldsymbol{r}^2\right)^{\gamma}}{4}
\ln\left(\frac{1}{\Lambda|\boldsymbol{r}|}+e\right)
\right]~.
\label{eq::S0}
\end{equation}
We use $\gamma=1.118$ and $\Lambda=0.24$~GeV, as determined from a fit to the inclusive DIS structure functions measured at HERA~\cite{Albacete:2010sy}. For the proton target, the fit gives an initial saturation scale $Q_{s0}^2=0.16~\mathrm{GeV}^2$, while the effective transverse area is fixed by $\pi R_p^2=\sigma_0/2$, with $\sigma_0=33.105$~mb. To reveal the saturation effects, we also perform the calculation for a gold nucleus with atomic number $A=197$. In this case, we take the initial nuclear saturation scale to lie in the range $3Q_{s0}^2 < Q_{s0A}^2 < 5Q_{s0}^2$ for estimates. The effective transverse area of the nucleus is assumed to be related to that of the proton by~\cite{Marquet:2025jdr}
\begin{align}
	\pi R_{A}^2=A \left(\frac{Q_{s 0, p}^2}{Q_{s 0, A}^2}\right)^{\gamma}\pi R_{p}^2,
\end{align}
which ensures that the nuclear modification factor of the energy pattern, $\frac{1}{A} \frac{d \Sigma_{e A}}{d \Sigma_{e p}}$, tends to unity in the dilute limit, where $\theta Q\gg Q_{s}$.

 \begin{figure}[!t]
\centering
\includegraphics[width=0.4\linewidth]{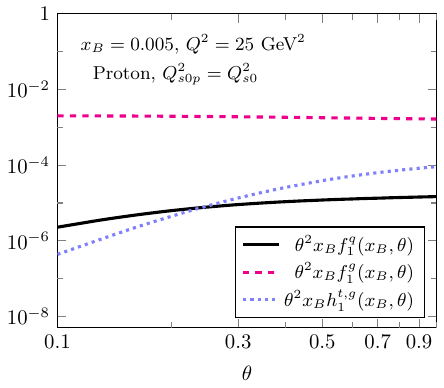}
\includegraphics[width=0.4\linewidth]{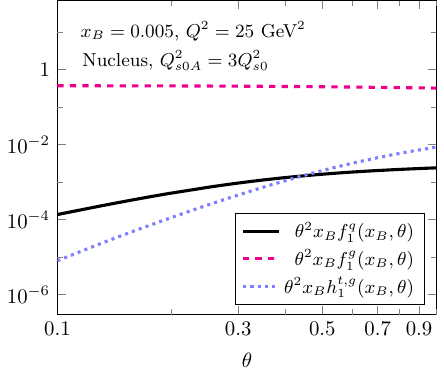}
\caption{Predictions for the unpolarized quark NEC $f_1^q$, the unpolarized gluon NEC $f_1^g$, and the linearly polarized gluon NEC $h_1^{t,g}$ of the proton (left panel) and the gold nucleus (right panel).}
\label{fig:f_A1}
\end{figure}

 \begin{figure}[!t]
\centering

\includegraphics[width=0.43\linewidth]{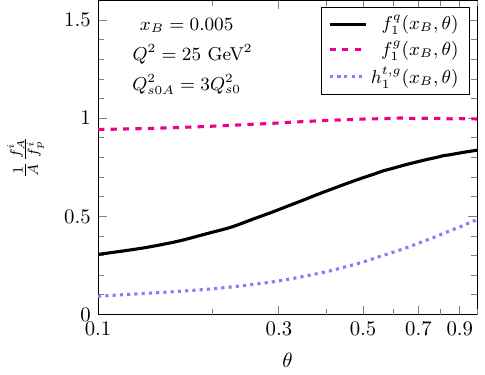}
\caption{Nuclear modification factors of the unpolarized quark NEC, the unpolarized and the linearly polarized gluon NEC as functions of $\theta$.}
\label{fig:f_A2}
\end{figure}

In Fig.~\ref{fig:f_A1}, we show the CGC predictions for the unpolarized gluon NEC $f_1^g$ and the linearly polarized gluon NEC $h_1^{t,g}$ as functions of $\theta$ for the proton and gold targets. For comparison, we also include the unpolarized quark NEC $f_1^q$ computed based on Ref.~\cite{Mantysaari:2025mht}. To make the $\theta$-angular dependence more transparent, we plot the distributions multiplied by $\theta^2$. After this rescaling, we find that $f_1^g$ is nearly flat and remains a few orders of magnitude larger than both $f_1^q$ and $h_1^{t,g}$ across the plotted range. By contrast, $\theta^2 h_1^{t,g}$ and $\theta^2 f_1^q$ show similar $\theta$-dependence and both increase with $\theta$. In the large-$\theta$ region, where $\theta Q\gg Q_s$, $\theta^2 f_1^q$ and $\theta^2h_1^{t,g}$ tend to approach constant values, consistent with the $1/\theta^2$ behavior obtained from the collinear expansion in the dilute limit,  as discussed in Sec.~\ref{sec:dilute_limit}. Although $\theta^2 h_1^{t, g}$ continues to increase, its growth gradually slows down, indicating an approach to a constant at sufficiently large $\theta$.

To illustrate the sensitivity to the saturation scale, Fig.~\ref{fig:f_A2} shows the nuclear modification factor $\frac{1}{A}\frac{f_A^i}{f_p^i}$,
comparing a proton target with a nuclear target for which $Q_{s0A}^2=3Q_{s0p}^2$. We find almost no nuclear suppression for the unpolarized gluon NEC $f_1^g$ across the full $\theta$ range, which is consistent with the results of the unpolarized gluon jet fracture function presented in Ref.~\cite{Caucal:2025qjg}. By contrast, the suppression for the linearly polarized gluon NEC $h_1^{t, g}$ and the quark NEC $f_1^q$ is much stronger in the small-$\theta$ region and becomes weaker at larger $\theta$. This shows that the latter two distributions are considerably more sensitive to saturation effects. In particular, the sensitivity to the linearly polarized gluon NEC $h_1^{t, g}$ is significantly more pronounced than to the unpolarized quark NEC $f_1^q$. This is expected, because the gluon NECs originate from an adjoint dipole S-matrix, while the quark NEC is yielded by the fundamental dipole.

\begin{figure}[!t]
\centering

\includegraphics[width=0.4\linewidth]{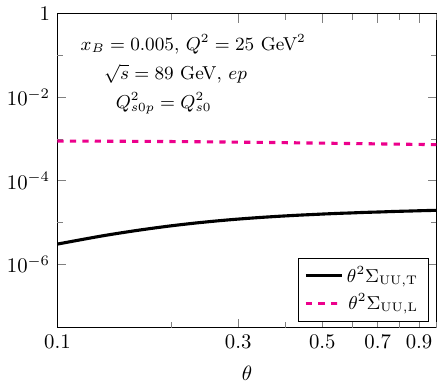}
\includegraphics[width=0.4\linewidth]{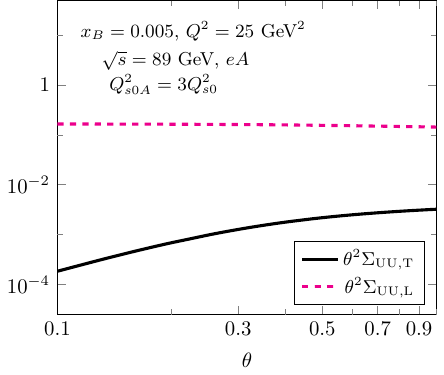}

\caption{Predictions for the DIS energy-pattern structure functions $\Sigma_{UU,T}$ and $\Sigma_{UU,L}$ as functions of $\theta$ in $ep$ collisions (left panel) and $eA$ collisions (right panel).}
\label{fig:UU_A1}
\end{figure}

\begin{figure}[!t]
\centering
\includegraphics[width=0.43\linewidth]{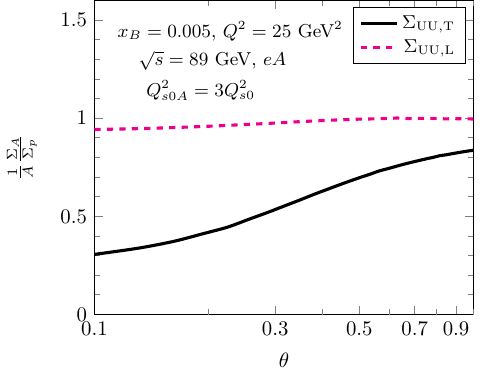}

\caption{Nuclear modification factors for the DIS energy-pattern structure functions, defined as $\frac{1}{A}\frac{\Sigma_A}{\Sigma_p}$, as functions of $\theta$.}
\label{fig:UU_A2}
\end{figure}
Fig.~\ref{fig:UU_A1} shows the $\theta$-dependence of the azimuthally averaged structure functions $\Sigma_{UU,L}$ and $\Sigma_{UU,T}$ for both the proton and nuclear targets, while Fig.~\ref{fig:UU_A2} presents their nuclear modification factors, given as $\frac{1}{A}\frac{\Sigma_A}{\Sigma_p}$. We observe that the longitudinal-photon structure function $\Sigma_{UU,L}$ and its nuclear modification factor are nearly flat over almost the entire $\theta$ range, exhibiting a behavior similar to that of the unpolarized gluon NEC shown in Fig.~\ref{fig:f_A1}. Combined with Eq.~\eqref{eq:UUL}, this indicates that the unpolarized gluon NEC dominates over the quark NEC in $\Sigma_{UU,L}$ at small $x_B$. By contrast, the transverse-photon structure function $\Sigma_{UU,T}$, which provides the baseline for the $\cos 2\phi$ azimuthal asymmetry, is determined by the unpolarized quark NEC at leading order in $\alpha_s$. Accordingly, its behavior in Fig.~\ref{fig:UU_A1} closely follows that of the quark NEC shown in Fig.~\ref{fig:f_A1}.

\begin{figure}[!t]
\includegraphics[width=0.434\linewidth]{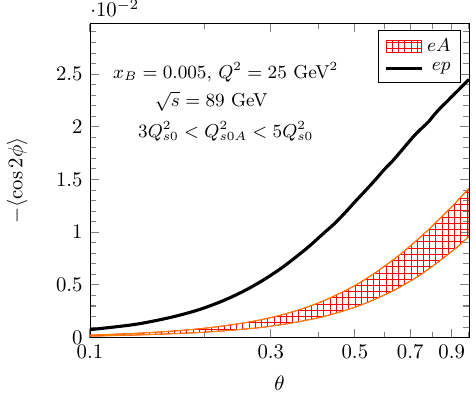}
\includegraphics[width=0.44\linewidth]{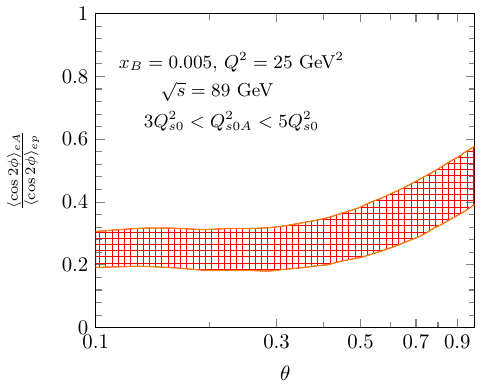}
\caption{Predictions for the azimuthal asymmetry $\langle \cos 2\phi \rangle$ (left panel) as a function of $\theta$ as well as its nuclear modification factor (right panel) between $eA$ and $ep$ collisions.
}
\label{fig:cos2phi_A}
\end{figure}

Fig.~\ref{fig:cos2phi_A} shows the predicted azimuthal asymmetry $\langle \cos 2\phi \rangle$ as a function of $\theta$, together with the corresponding nuclear modification factor between $eA$ and $ep$ collisions at the EIC. From the left panel, we see that $\langle \cos 2\phi \rangle$ is negative and increases in magnitude with $\theta$ for both targets. We also find that the asymmetry is systematically smaller in $eA$ collisions than in $ep$ collisions. To quantify the nuclear effect, we follow Ref.~\cite{Tong:2022zwp} and define the nuclear modification factor for the azimuthal asymmetry as
\begin{align}
    {\cal R}_{eA} = \frac{\langle \cos 2\phi \rangle_{A}}{\langle \cos 2\phi \rangle_{p}} \, .
\end{align}
As shown in the right panel of Fig.~\ref{fig:cos2phi_A}, ${\cal R}_{eA}$ stays well below unity throughout the plotted $\theta$ region, indicating a sizable nuclear suppression of the asymmetry. This suppression is most pronounced at small $\theta$ and becomes weaker toward larger $\theta$, in agreement with the expectation that saturation effects dominate when $\theta Q \lesssim Q_s$. This demonstrates that the $\cos 2\phi$ asymmetry in the DIS energy pattern is highly sensitive to the nuclear suppression of the linearly polarized gluon NEC, making it a promising probe of saturation effects at the future EIC.

\section{Summary}
\label{sec:conclusion}
In this work, we have evaluated leading-twist gluon fracture functions and NECs within the CGC formalism, starting from their operator definitions. Among the eight independent leading-twist gluon components, we find that only the unpolarized and linearly polarized components for an unpolarized target are nonvanishing in the strict eikonal limit. Both can be expressed entirely in terms of the adjoint dipole $S$-matrix, as detailed in Sec.~\ref{sec:CGC_results}. Together with previous results in the sea-quark sector, our findings in the gluon sector establish a unified framework for probing small-$x$ nucleon tomography in the TFR of DIS at eikonal accuracy.

We have further proposed the linearly polarized gluon NEC $h_{1}^{t,g}$ as a novel and sensitive probe of saturation effects through the $\cos 2\phi$ azimuthal asymmetry of the DIS energy pattern in the TFR. As shown in Eq.~\eqref{eq:cos2phi_num}, this asymmetry is governed by the ratio of the linearly polarized gluon NEC $h_{1}^{t,g}$ to the unpolarized quark NEC $f_{1}^{q}$. Within the CGC framework, these two NECs are controlled by the adjoint and fundamental dipole amplitudes, respectively. Our numerical analysis for EIC kinematics shows that the magnitude of this $\cos 2\phi$ asymmetry is substantially reduced in $e+A$ relative to $e+p$ scattering. The nuclear suppression, quantified by the modification factor ${\cal R}_{eA}$, is strongest at small $\theta$ and gradually weakens toward the dilute region. The proposed observable therefore provides a distinctive nuclear signature of gluon saturation.

Because the proposed NEC-induced asymmetry is formulated within collinear factorization, it is not subject to the Sudakov suppression or soft-radiation-induced $\cos 2\phi$ contamination encountered in TMD-based dijet observables. Its measurement requires only a single inclusive energy flow rather than the reconstruction of a dijet system. These features make the TFR energy-pattern asymmetry a theoretically clean and experimentally attractive probe of linearly polarized gluons and the onset of saturation at the future EIC.

To formulate the $\cos 2\phi$ energy-pattern asymmetry, we have first taken the TFR limit, where the relevant structure function $\Sigma_{UU}^{\cos 2\phi}$ factorizes in terms of the gluonic NEC $h_{1}^{t,g}$, and then the small-$x_B$ limit, where $h_{1}^{t,g}$ further factorizes in terms of dipole amplitudes. To fully establish the consistency between CGC and collinear factorization in the TFR, it would also be important to derive the result directly from the full CGC cross section and subsequently take the TFR limit, following, for example, the approach in Ref.~\cite{Caucal:2025qjg}.
It would be possible to extend our analysis to include subeikonal corrections to fracture functions and NECs, using the techniques developed for the TMDs~\cite{Kovchegov:2015pbl,Kovchegov:2018znm,Kovchegov:2018zeq}. Such corrections introduce helicity-dependent and $C$-odd operator structures that may generate target-spin-dependent gluon NECs and fracture functions, giving rise to novel spin and azimuthal asymmetries that vanish at eikonal accuracy. Moreover, while this work has focused on Weizsäcker--Williams-type gluon NECs and fracture functions relevant to DIS, it would be important to investigate alternative gauge-link topologies and clarify the process dependence and universality properties of the resulting distributions. These developments would extend the applicability of gluonic NECs and fracture functions to polarized observables and hadronic collisions in the saturation regime.

\acknowledgments  We thank Feng Yuan, Jian Zhou, Paul Caucal and Farid Salazar for helpful discussions. This work is supported by the Research Council of Finland, the Centre of Excellence in Quark Matter and projects 338263 and 359902, and by the European Research Council (ERC, grant agreements No. ERC-2023-101123801 GlueSatLight and No. ERC-2018-ADG-835105 YoctoLHC). Y.T. was supported by the National Research Council of Thailand
(NRCT) via the project number 220677 (contract number N42A690266), by the National Science, Research and Innovation Fund (NSRF) via the Program Management Unit for Human Resources \& Institutional Development, Research and Innovation (grant number B39G680009), and by grants for development of new faculty members, Ratchadaphiseksomphot Fund, Chulalongkorn University.
The content of this article does not reflect the official opinion of the
European Union and responsibility for the information and views expressed therein lies entirely with the
authors.

\appendix

 \section{Gluonic NECs and fracture functions from gluon collinear splittings}
  \label{appendix:collinear}

In this appendix, we derive the gluonic fracture functions at large-$\bs p_{g\perp}$ and the gluonic NECs at large-$\theta$ from gluon collinear splittings within the collinear formalism. In particular, we work in the target light-cone gauge
\begin{align}
  n\cdot A=A^+=0~,
\end{align} and the associated polarization sum of gluons is given as:
\begin{align}
\sum_{\lambda=\pm1} \epsilon_\sigma^*(p_g,\lambda) \epsilon_{\sigma'}(p_g,\lambda)=-g_{\sigma\sigma'}+\frac{ n_\sigma p_{g\sigma'}+ n_{\sigma'} p_{g\sigma}}{ n \cdot p_g}=\tilde \Pi_{\sigma\sigma'}(p_g)~.
\end{align}
In this gauge, only one diagram contributes in the gluonic channel. For simplicity, we consider the gluonic fracture function for producing a gluon with the momentum $p_g$. The contribution is written as
\begin{align}
  {\cal M}^{\alpha \beta} _{G,\text{FrF}}=&\frac{1}{xP^+}\int dk_g^+\delta(x P^++p_g^+-k_g^+)(-i)\big[ xP^+ g_\perp^{\alpha\mu'}-(k_g-p_g)^\alpha_\perp n^{\mu'}\big]
  i\big[ xP^+ g_\perp^{\beta\nu'}-(k_g-p_g)^\beta_\perp n^{\nu'}\big]
  \notag \\
  &
  \frac{i\tilde \Pi_{\mu'\mu}(k_g-p_g)}{(k_g-p_g)^2}  \frac{-i\tilde \Pi_{\nu\nu'}(k_g-p_g)}{(k_g-p_g)^2}
(-g_sf^{abc})  V^{ \rho\mu\sigma}(k_{g},p_g-k_{g},-p_g)
\notag \\
&\times
(-g_sf^{\tilde acb})  V^{ \tilde \rho \tilde \sigma\nu}(k_{g},-p_g,p_g-k_{g})\tilde \Pi_{\sigma\tilde \sigma}(p_g)  \notag \\
&\times \int \frac{dy^-}{(2\pi)}e^{i k_g^+ y^-}\langle P|A^a_{\perp\rho}(y) A_{\perp\tilde\rho}^{\tilde a}(0) |P\rangle~,
\end{align}
where we have made the collinear approximation for the incoming gluon, $k_g^\mu\approx (x_g P^+,0, \bs 0_\perp)$. The tensor structure of the three-gluon vertex is
\begin{align}
   V^{\alpha\beta\gamma}(p,q,r) =(p-q)^\gamma g^{\alpha\beta}+(q-r)^\alpha g^{\beta\gamma}+(r-p)^\beta g^{\alpha\gamma}.
\end{align}
The collinear gluon distribution is obtained by
\begin{align}
 \int \frac{dy^-}{2\pi}e^{i k_g^+ y^-}\langle P|A^a_{\perp\rho}(y) A_{\perp\tilde \rho}^{\tilde a}(0) |P\rangle= \frac{-g_{\perp\rho \tilde \rho}}{2} \frac{\delta_{a\tilde a}}{N_c^2-1}\frac{1}{x_g P^+} f_g(x_g)~.
\end{align}

\paragraph{Unpolarized gluon case}The unpolarized gluonic fracture function is projected by
\begin{align}
       u_{1}^g=&\big(-g^{\alpha\beta}_\perp\big){\cal M}_{G,\text{FrF},\alpha \beta}~.
\end{align}
Thus, we obtain
\begin{align}
  u_{1}^g(x,\xi,\bs p_{g\perp}^2)=\frac{4g_s^2 N_c}{\bs p_{g\perp}^2} \frac{(\xi^2+x^2+\xi x)^2}{x(\xi+x)^4} x_g f(x_g)~,
  \label{eq:ugcoll}
\end{align}
where $x_g= \xi+x$~. The associated NEC is given by
 \begin{align}
f_{1}^g(x,\theta)=&\sum_{a}\frac{1}{2}
\int_0^{1-x}d \xi_a\xi_a\Big(\frac{\xi_a P^+}{\sqrt{2}}\Big)^2 \frac{1}{(2\pi)^32\xi_a}u_{1}^g
(x,\xi_a,\bs p_{a\perp}^2)
\Big \vert_{\bs p_{a\perp}=\frac{\theta\xi_a P^+}{\sqrt{2}}
 (\cos \phi, \sin \phi)}~.
\end{align}
Changing the variable from $\xi$ to $z$ by $\xi=\frac{1-z}{z}x$, we obtain the gluon NEC in the collinear limit:
\begin{align}
f_{1}^g(x, & \left.\theta\right)=\frac{1}{2\pi}\frac{\alpha_s}{2 \pi} \frac{1}{\theta^2} \int_x^1 \frac{d z}{z} \frac{x(1-z)}{z} 2 C_A \frac{(1-z+z^2)^2}{z(1-z)}f_g\left(\frac{x}{z}\right)~.
\label{app:f1}
\end{align}

\paragraph{Linearly polarized gluon case}The linearly polarized gluonic fracture function is projected by
\begin{align}
\frac{t_{1}^{h,g}}{2M^2} =&\frac{2}{ (\bs p_{g\perp}^2)^2}\Big( p_{g\perp}^\alpha   p_{g\perp}^\beta + \frac{1}{2}g_\perp^{\alpha \beta} \bs p_{g\perp}^2\Big){\cal M}_{G,\text{FrF},\alpha \beta}~.\end{align}
Then, we obtain
\begin{align}
\frac{t_{1}^{h,g}(x,\xi,\bs p_{g\perp}^2)}{2M^2}=\frac{4g_s^2 C_A}{(\bs p_{g\perp}^2)^2} \frac{\xi^2}{x(\xi+x)^2}  x_gf(x_g)~,
  \label{eq:tgcoll}
\end{align}
where $x_g= \xi+x$~.
 The associated NEC is given by
 \begin{align}
   h_{1}^{t,g}(x,\theta)=&\sum_{a}\frac{1}{2}
\int_0^{1-x}d \xi_a \xi_a\Big(\frac{\xi_a P^+}{\sqrt{2}}\Big)^2  \Big( \frac{\theta \xi_a P^+}{\sqrt{2}}\Big)^2\frac{1}{2\xi_a(2\pi)^3}\frac{t_{1}^{h,g}(x,\xi_a,\bs p_{a\perp}^2)}{2M^2}
\Big \vert_{\bs p_{a\perp}=\frac{\theta\xi_a P^+}{\sqrt{2}}
 (\cos \phi, \sin \phi)}~.
\end{align}
Changing the variable from $\xi$ to $z$ by $\xi=\frac{1-z}{z}x$, we obtain
\begin{align}
  h_{1}^{t,g}(x,\theta)
  =&\frac{1}{2\pi}\frac{\alpha_s}{2 \pi} \frac{1}{\theta^2}\int^1_x \frac{dz}{z}
\frac{2C_A(1-z)}{z} (1-z)\frac{x}{z}f_g(\frac{x}{z})~.
\label{app:h1}
\end{align}
The expressions in Eq.~\eqref{app:f1} and Eq.~\eqref{app:h1} are consistent with the results, as given in Ref.~\cite{Guo:2024jch}.

\bibliographystyle{JHEP}
\bibliography{ref.bib}
\end{document}